\documentclass[12pt]{article}
\usepackage[utf8]{inputenc} 
\usepackage[T1]{fontenc}    
\usepackage{url}            
\usepackage[breaklinks=true]{hyperref}

\usepackage{booktabs}       
\usepackage{amsfonts}       
\usepackage{nicefrac}       
\usepackage{microtype}      
\usepackage{amsmath}
\usepackage{float}
\usepackage{graphicx}
\usepackage{caption}
\usepackage{subcaption}
\usepackage{relsize}
\usepackage{mathtools}
\usepackage{multirow}
\usepackage[toc,page]{appendix}
\usepackage{breakcites}

\newcommand{\thickhline[1]}{%
    \noalign {\ifnum 0=`}\fi \hrule height #1
    \futurelet \reserved@a \@xhline
}

\usepackage[table]{xcolor}
\usepackage{array}

\DeclareFontFamily{OT1}{mathc}{}
\DeclareFontShape{OT1}{mathc}{m}{it}{<-> mathc10}{}
\DeclareMathAlphabet{\mathabxcal}{OT1}{mathc}{m}{it}

\hypersetup{
    colorlinks=true,
    citecolor=black,
    linkcolor=black,     
    urlcolor=blue
    }

\newcommand{\blind}{0}

\begin{document}

\def\spacingset#1{\renewcommand{\baselinestretch}%
{#1}\small\normalsize} \spacingset{1}


\if0\blind
{
  \title{\bf Bayesian spatio--temporal disaggregation modeling using a diffusion-SPDE approach: a case study of Aerosol Optical Depth in India}
  \author{\textbf{Fernando Rodriguez Avellaneda} \\
Computer, Electrical and Mathematical Sciences and Engineering Division,\\
King Abdullah University of Science and Technology (KAUST),\\
Thuwal, Saudi Arabia,\\
    \textbf{Paula Moraga} \\
Computer, Electrical and Mathematical Sciences and Engineering Division,\\
King Abdullah University of Science and Technology (KAUST),\\
Thuwal, Saudi Arabia}
  \maketitle
} \fi

\if1\blind
{
  \bigskip
  \bigskip
  \bigskip
  \begin{center}
    {\LARGE\bf Bayesian spatio--temporal disaggregation modeling using a diffusion-SPDE approach: a case study of Aerosol Optical Depth in India}
\end{center}
  \medskip
} \fi

\bigskip
\begin{abstract}

Accurate estimation of Aerosol Optical Depth (AOD) is crucial for understanding climate change and its impacts on public health, as aerosols are a measure of air quality conditions.
AOD is usually retrieved from satellite imagery at coarse spatial and temporal resolutions. However, producing high-resolution AOD estimates in both space and time can better support evidence-based policies and interventions.
We propose a spatio-temporal disaggregation model
that assumes a latent spatio--temporal continuous Gaussian process observed through aggregated measurements. 
The model links discrete observations to the continuous domain and accommodates covariates to improve explanatory power and interpretability.
The approach employs Gaussian processes with separable or non-separable covariance structures derived from a diffusion-based spatio-temporal stochastic partial differential equation (SPDE). Bayesian inference is conducted using the INLA-SPDE framework for computational efficiency. Simulation studies and an application to nowcasting AOD at 550 nm in India demonstrate the model's effectiveness, improving spatial resolution from $0.75^\circ$ to $0.25^\circ$ and temporal resolution from 3 hours to 1 hour.

\end{abstract}

Keywords: 
Air pollution;
Disaggregation;
Non--separable covariance;
Spatio--temporal Modeling;
Difussion SPDE;
INLA
\vfill

\newpage
\spacingset{1.45} 
\section{Introduction}
\label{introduction}

Aerosols, commonly quantified through Aerosol Optical Depth (AOD), are a fundamental component of atmospheric composition with wide-ranging implications for both climate and human health \cite{rapetal2013aod}. In particular, AOD at 550 nanometers (nm) is a dimensionless quantity that measures the degree to which aerosols (e.g., dust, smoke, haze) scatter or absorb light in a vertical column of the atmosphere, using the middle of the visible spectrum (550 nm) as the reference wavelength. AOD is commonly combined with chemical transport models and ground observations to estimate and forecast ground--level particulate matter, thereby supporting air quality monitoring \cite{yu2023enviromentalAOD}. Elevated AOD measurements are frequently associated with high risks of respiratory and cardiovascular conditions that could include asthma, chronic obstructive pulmonary disease (COPD), and ischemic heart disease \cite{wang2013acute}. Beyond health effects, aerosols also modify the Earth's radiative balance, cloud microphysics, and precipitation patterns, thereby linking AOD to broader climate processes with negative impacts on ecosystems and agricultural productivity \cite{tai2014threat}.

Understanding the spatio--temporal distribution of AOD is critical for informing policies aimed at reducing its adverse effects on human health, ecosystems, and food security. A significant obstacle, however, lies in the limited availability of high–-resolution AOD data; ground monitoring stations remain sparse in many parts of the world, while reanalysis products such as CAMS \cite{inness2019cams} or satellite retrievals from instruments such as MODIS \cite{remeretal2005modis} and MISR \cite{kahnetal2005misr} provide global coverage only at coarse spatial (tens of kilometers) and temporal (3–-6 hours) resolutions, which are often insufficient for public health and environmental applications where fine-scale variability is crucial \cite{vanDonkelaar2019review,shaddickEetall2018globalpm25}. This mismatch between the spatial resolution of available data and the finer scales required is often referred as the change of support problem \cite{patiletal2024disaggregation,avellaneda2025multivariatedisaggregationmodelingair}.

In this study, we develop a spatio-temporal modeling framework that enable the disaggregation of coarsely aggregated AOD data to produce high--resolution estimates in both space and time.
We evaluate the performance of our model by downscaling AOD at 550 nm in India, a region charaterized by high aerosol loading, diverse topography, and significant public health relevance.
Specifically, we use the Atmospheric Composition Reanalysis 4 (EAC4) dataset from the European Centre for Medium-Range Weather Forecasts (ECMWF) \cite{inness2019cams}. The dataset provides AOD values on a spatial grid of $0.75^\circ \times 0.75^\circ$ in latitude and longitude with a temporal resolution of 3 hours. In this study, we perform spatio-temporal disaggregation to achieve a finer spatial resolution of $0.25^\circ$ and 1 hour in time, thereby providing a more detailed representation of aerosol variability that can support improved analyses of its impacts on air quality and health.

The problem of disaggregation has been widely studied in the spatial framework, particularly in the context of air pollution \cite{patiletal2024disaggregation}. Existing approaches can be broadly classified into geostatistical methods and machine learning models. Machine learning methods are often computationally efficient and can handle large datasets, but they usually treat covariates in a black-box manner and operate at aggregated levels, which limits interpretability and their ability to capture spatial dependence explicitly \cite{monteiroetal2019dismachinelearning}. In contrast, geostatistical approaches leverage spatial autocorrelation to improve predictions and allow the explicit inclusion of covariates through Gaussian process models \cite{tanaka2019spatiallya,yousefi2019multitaska}. However, these models often become computationally prohibitive, as likelihood-based inference and prediction typically require costly approximations due to the lack of closed-form solutions \cite{banerjee2013efficientgp,heaton2019casestudycompetition}.

Bayesian melding and downscaler models have been presented by several authors to combine spatially misaligned data \cite{zhongandmoraga24}. For example, a Bayesian spatio-temporal downscaler was proposed by \cite{berrocal2010aspatiotemporal}, linking coarse numerical model outputs with point observations through spatially varying coefficients driven by Gaussian processes.
\cite{moraga2017geostatistical} introduced a Bayesian melding that assumes an underlying continuous latent Gaussian Random Field (GRF) observed at an aggregated level and uses INLA and SPDE for fast Bayesian inference \cite{lindgrenandrue15}, applied to map air pollution in California, USA.
Building on this idea, \cite{avellaneda2025multivariatedisaggregationmodelingair} extended the approach to a multivariate setting, applying it in Portugal and Italy to disaggregate PM2.5, PM10, and Ozone concentrations at finer spatial resolutions. 

More recently, scalable developments such as \texttt{DALIA} \cite{gaedkemerzhäuser2025acceleratedspatiotemporalbayesianmodeling} have addressed the problem of spatial disaggregation by utilizing efficient Bayesian inference for spatio-temporal multivariate Gaussian processes, thereby overcoming the computational bottlenecks of high-dimensional models through sparse precision formulations and parallelization strategies. While this approach used multivariate spatio-temporal data, its disaggregation is restricted to the spatial domain.
In parallel, machine learning approaches have introduced attention-based models to disaggregate spatio--temporal data, typically from coarse administrative units (e.g., census tracts) to finer ones (e.g., city blocks). A notable example is the Structurally-Aware Recurrent Network (SARN), which integrates spatial attention layers with recurrent neural networks to jointly capture spatial interactions and temporal dependencies \cite{han2024sarnstructurallyawarerecurrentnetwork}. 

Previous models have mainly focused on spatial disaggregation, maintaining computational scalability, and to the best of our knowledge none have explicitly addressed the joint spatio--temporal disaggregation problem.
In this paper, we propose a spatio--temporal disaggregation model for normally distributed data, applied here to aerosol optical depth (AOD) at 550 nm but broadly applicable to other spatio–-temporal contexts. Our model assumes the existence of a spatio--temporal latent Gaussian process with either a separable or non--separable covariance structure in space and time, providing the flexibility to capture realistic dependencies in space and time. Computational efficiency is achieved through the SPDE--GMRF framework, with inference performed using the \texttt{INLA} and \texttt{INLAspacetime} packages \cite{rue_et_al_r_inla_2009,lindgren2023inlaspacetime}. This approach directly addresses the challenge of linking data observed at aggregated levels to an underlying continuous latent process, enabling predictions at finer spatial and temporal resolutions.

The remainder of the paper is organized as follows. Section~\ref{sec:methodology} presents the methodological framework for the spatio-temporal disaggregation model, and the inference strategy adopted in this work. Section~\ref{sec:sim} reports the results of a simulation study evaluating the performance of our spatio--temporal disaggregation model. Section~\ref{sec:app} applies the model to AOD data in India, incorporating altitude as a covariate. Finally, Section~\ref{sec:con} concludes with a discussion and outlines directions for future research.

For reproducibility, the code used to implement this approach is publicly available at the GitHub repository \url{https://github.com/fravellaneda/spatio-temporal-disaggregation}.
The repository contains documentation that allows researchers replicate our analyses and use the methodology in their own applications.

\section{Methodology}
\label{sec:methodology}

In this section, we present the theoretical framework of our spatio--temporal disaggregation approach. First, we introduce the disaggregation model, which assumes the existence of a continuous latent process observed on aggregated supports.
Second, we specify the structure of the latent random effect via a diffusion SPDE, which leads to a Gaussian process that may have a separable or non--separable covariance structure. 
Finally, we describe the inference strategy based on the INLA-SPDE formulation, which was implemented using the \texttt{R-INLA} \cite{rue_et_al_r_inla_2009} and \texttt{INLAspacetime} packages \cite{lindgren2023inlaspacetime}.

\subsection{Spatio-temporal disaggregation model}

Let $Y_{ij}$ denote the set of observations in regions $i=1,\ldots,N$ and time points $j=1,\ldots,M$. We assume that underlying all observations there is a spatio--temporal continuous latent field that is observed in the aggregated resolution. This latent continous model can be expressed as the combination of some fixed effects and random effects as
\begin{align}
\begin{split}
    y(\mathbf{s},t) 
        &= \; \beta\,X(\mathbf{s},t) \;+\; z(\mathbf{s},t) \\[6pt]
    &= \;
       \begin{bmatrix}
           \beta_{0} \\
           \beta_{1} \\
           \vdots \\
           \beta_{p}
       \end{bmatrix}
       \cdot
       \big[1,\,x_{1}(\mathbf{s},t),\,\ldots,\,x_{p}(\mathbf{s},t)\big]
       \;+\; z(\mathbf{s},t),
\end{split}
\label{equ:cont-1}
\end{align}
where $\beta\,X(\mathbf{s},t)$ represents the $p$ spatio--temporal fixed effects (including the intercept), and $z(\mathbf{s},t)$ denotes a spatio-temporal Gaussian process.


From the continuous spatio--temporal process $y(\mathbf{s},t)$, we define an aggregated process over a spatial region $R \subseteq \mathbb{R}^2$ and a temporal window $T \subseteq [0,\infty)$ as
\begin{equation}
    u(R,T) = \frac{1}{|R|\cdot|T|}\int_{R} \int_{T} z(\mathbf{s},t)\,\mathrm{d}t\,\mathrm{d}\mathbf{s},
    \label{equ:cont-2}
\end{equation}
with $|R| = \int_{R} \mathrm{d}\mathbf{s}$ and $|T| = \int_{T} \mathrm{d}t$. 
We then assume that the observed spatio--temporal values arise from the aggregation of the continuous latent process 
together with a measurement error term. 
In this case, Equation~\ref{equ:cont-1} aggregates to the discrete process
\begin{equation}
    y(R_i,T_j) = \beta\,X(R_i,T_j) + z(R_i,T_j) + v_{ij},
    \label{equ:areal}
\end{equation}
where $X(R_i,T_j)$ represents the average covariates over the spatial region $R_i$ and time period $T_j$, $z(R_i,T_j)$ is the aggregated process defined in Equation~\ref{equ:cont-2},  and $v_{ij}$ is an unstructured spatio--temporal noise term, modeled as $v_{ij} \sim \mathcal{N}(0, \sigma^2_{\epsilon})$.

\subsection{Diffusion model}
\label{sec:diffusion}

The spatio-temporal structure of the random process $z(\mathbf{s},t)$ in Equation~\ref{equ:cont-1} is modeled as a Gaussian process that arises as the solution of a diffusion-based spatio--temporal stochastic partial differential equation (SPDE) \cite{lindgren2023inlaspacetime}. This equation comes from the operator $L_s = \gamma_s^2 - \Delta$, where $\gamma_s > 0$, $\Delta$ denotes the Laplacian, and the domain $\Omega \subseteq \mathbb{R}^d$ is equipped with suitable boundary conditions to ensure well-posedness on compact subsets. Based on this operator, the precision matrix for the generalized Whittle-Matérn covariance class could be defined as $ Q(\gamma_s, \gamma_e, \alpha) = \gamma_e^2 L_s^{\alpha}$ and the spatial Gaussian random field $\varphi(\mathbf{s})$ satisfies the spatial SPDE
\begin{equation*}
\gamma_e L_s^{\alpha/2} \varphi(\mathbf{s}) = \mathcal{W}(\mathbf{s}), \quad \mathbf{s} \in \Omega,
\end{equation*}
where $\mathcal{W}(\mathbf{s})$ denotes spatial white noise. From this, we define a spatio--temporal noise process $\mathrm{d} \mathcal{E}_Q(\mathbf{s}, t)$ as temporally uncorrelated but spatially correlated Gaussian noise, with precision operator $Q = Q(\gamma_s, \gamma_e, \alpha_e)$ for a non-negative parameter $\alpha_e$.

Using this noise formulation, we specify a spatio-temporal diffusion model with three non-negative smoothness parameters $(\alpha_t, \alpha_s, \alpha_e)$ and three positive scale parameters $(\gamma_t, \gamma_s, \gamma_e)$. The resulting spatio--temporal SPDE is given by
\begin{equation}
\left(\gamma_t \frac{d}{dt} + L_s^{\alpha_s} \right)^{\alpha_t} v(\mathbf{s}, t) = \mathrm{d} \mathcal{E}_Q(\mathbf{s}, t), \quad (\mathbf{s}, t) \in \Omega \times \mathbb{R}.
\label{equ:diffusion}
\end{equation}
The solution $v(\mathbf{s}, t)$ to Equation~\eqref{equ:diffusion} defines a Gaussian process with a specific covariance structure determined by its parameters $(\alpha_t, \alpha_s, \alpha_e)$. In this setting, the parameter $\alpha_e$ controls the degree of separability: setting $\alpha_e = 0$ results in a fully nonseparable model. To further characterize the separability structure, \cite{lindgren2023inlaspacetime} introduces a nonseparability parameter.
\begin{equation*}
    \beta_s = 1 - \frac{\alpha_e}{\alpha} = 1 - \frac{\alpha_e}{v_s + d/2},
\end{equation*}
where $\nu_s$ denotes the smoothness of the spatial process, and $\beta_s \in [0,1]$ quantifies the degree of separability. The values $\beta_s = 0$ and $\beta_s = 1$ correspond to fully separable and fully nonseparable models, respectively. Notably, a special case arises when $\alpha_s = 0$, which implies that $\nu_s = \alpha_e - d/2$, and therefore $\beta_s = 0$, resulting in a separable model. Specifically, in the simulation study, we used a separable model with parameters $(\alpha_t, \alpha_s, \alpha_e) = (1, 0, 2)$ and a non--separable model with parameters $(\alpha_t, \alpha_s, \alpha_e) = (1, 2, 1)$. Both models share the same spatial and temporal smoothing parameters, with $\nu_s = 1$ and $\nu_t = 1/2$.

To improve interpretability and comparability across different model specifications, we reparameterize the scale parameters $(\gamma_t, \gamma_s, \gamma_e)$ following the approach proposed in \cite{lindgren2023inlaspacetime}. This transformation for the separable cases relates the original scale parameters to physically meaningful quantities: the marginal variance $\sigma^2$, the spatial correlation range $r_s$, and the temporal correlation range $r_t$. The relationships are defined as follows:

\begin{align}
\begin{split}
\sigma^2 & = \frac{C_{\mathbb{R}, \alpha_t} \, C_{\mathbb{R}^d, \alpha}}{\gamma_t \, \gamma_e^2 \, \gamma_s^{2\alpha - d}}, \\
r_s & = \gamma_s^{-1} \sqrt{8 \nu_s}, \\
r_t & = \gamma_t \, \gamma_s^{-\alpha_s} \sqrt{8\left(\alpha_t - \frac{1}{2}\right)},
\end{split}
\label{equ:trans-scale}
\end{align}

where the constants

\begin{equation*}
C_{\mathbb{R}^d, \alpha} = \frac{\Gamma(\alpha - d/2)}{\Gamma(\alpha)(4\pi)^{d/2}} \quad \text{and} \quad \alpha = \alpha_e \left(\alpha_t - \frac{1}{2} \right).
\end{equation*}

In the non--separable setting, the parameters $\sigma^2$ and $r_s$ retain their usual interpretations as the marginal variance and spatial correlation range, respectively, consistent with the Matérn covariance structure described in \cite{Lindgren2011}. In contrast, the temporal range parameter $r_t$ reflects the temporal correlation of the spatial eigenfunction associated with the smallest eigenvalue of the Laplacian, that is, a spatially constant mode that evolves over time \cite{lindgren2023inlaspacetime}.

\subsection{Inference using INLA}

We aim to fit the spatio--temporal model defined in Equation~\ref{equ:cont-1} where the random effect $z(s,t)$ is the solution of a specific SPDE model defined in Section~\ref{sec:diffusion}. To this end, we use the connection between Gaussian processes and Gaussian Markov random fields (GMRFs) through the SPDE approach \cite{Lindgren2011}. In practice, we rely on the \texttt{R-INLA} and \texttt{INLAspacetime} packages, which provide an efficient and scalable framework for spatio–-temporal geostatistical modeling by approximating the GMRFs \cite{rue_et_al_r_inla_2009,lindgren2023inlaspacetime}.

The solution of the SPDE requires the construction of a triangulated mesh with $G$ nodes in space and a one dimensional temporal mesh with $D$ time points, yielding $G \times D$ basis functions in the joint spatio–-temporal domain. These basis functions, denoted by $\psi_{k,p}$, are piecewise linear functions that take the value $1$ at vertex $\mathbf{s}_k$ at time $t_j$, i.e., $\psi_{k,p}(\mathbf{s}_k,t_p)=1$, and $0$ otherwise. Using these basis functions, the latent process $z(\mathbf{s},t)$ in Equation~\ref{equ:cont-1} can be approximated as

\begin{equation*}
z(\mathbf{s},t) \approx \sum_{k=1}^{G} \sum_{p=1}^{D} \psi_{k,p}(\mathbf{s},t) \mathabxcal{z}_{k,p},
\end{equation*}
where $\mathabxcal{z}_{k,p}$ are the corresponding stochastic weights that approximate the GMRF in the spatio--temporal mesh nodes.

Now consider a spatial region $R_i \subseteq \mathbb{R}^2$, with $i=1,\ldots,N$, and a time interval $T_j$, with $j=1,\ldots,M$. Let $\{\mathbf{s}_k\}_{k=1}^{G_i}$ denote the spatial mesh points within region $R_i$, and $\{t_p\}_{p=1}^{D_j}$ the temporal mesh points within interval $T_j$. Extending the ideas of spatial disaggregation in \cite{moraga2017geostatistical} to the spatio--temporal setting, we can express the aggregated latent field $z(R_i,T_j)$ in terms of the spatio--temporal mesh as  

\begin{align}
    \begin{split}
        z(R_i,T_j) 
        &= \frac{1}{|R_i|\cdot|T_j|}\int_{R_i} \int_{T_j} z(\mathbf{s},t)\,\mathrm{d}t\,\mathrm{d}\mathbf{s} \\[0.5em]
        &\approx \frac{1}{|R_i|\cdot|T_j|}\int_{R_i} \int_{T_j} \sum_{k=1}^{G_i} \sum_{p=1}^{D_j} \psi_{k,p}(\mathbf{s},t)\,\mathabxcal{z}_{k,p}\,\mathrm{d}t\,\mathrm{d}\mathbf{s} \\[0.5em]
        &= \frac{1}{|R_i|\cdot|T_j|} \sum_{k=1}^{G_i} \sum_{p=1}^{D_j} \int_{R_i} \int_{T_j} \psi_{k,p}(\mathbf{s},t)\,\mathrm{d}t\,\mathrm{d}\mathbf{s}\,\mathabxcal{z}_{k,p} \\[0.5em]
        &\approx \frac{1}{|R_i|\cdot|T_j|} \sum_{k=1}^{G_i} \sum_{p=1}^{D_j} 
        \left(\int_{R_i}\int_{T_j} \psi_{k,p}(\mathbf{s}_k,t_p) 1_{\{\psi_{k,p}(\mathbf{s},t)\neq 0\}}\,\mathrm{d}t\,\mathrm{d}\mathbf{s}\right)\mathabxcal{z}_{k,p} \\[0.5em]
        &= \sum_{k=1}^{G_i} \sum_{p=1}^{D_j} \frac{|R_{ik}|\cdot|T_{jp}|}{|R_i|\cdot|T_j|}\,\mathabxcal{z}_{k,p} 
        = \sum_{k,p}^{G_i,D_j} A_{kp}^{ij}\,\mathabxcal{z}_{k,p},
    \end{split}
    \label{equ:connect}
\end{align}
where $|R_{ik}|$ is the portion of $R_i$ associated with vertex $\mathbf{s}_k$, $|T_{jp}|$ is the portion of $T_j$ associated with time point $t_p$, and $A_{kp}^{ij}$ are the entries of the spatio--temporal projection matrix linking the continuous latent field to the aggregated domain. 

In Equation~\ref{equ:connect}, the first approximation corresponds to the finite basis expansion used in the SPDE solution. The second approximation holds because the $(k,p)$-th basis function evaluates to one at its corresponding spatio--temporal mesh vertex, i.e., $\psi_{k,p}(\mathbf{s}_k,t_p)=1$. We assume that the number of unique spatial locations is $G$ and the number of unique time points is $D$, while the aggregated observations are available over $N$ regions and $M$ time intervals. Under this setting, the projection matrix $A$ is an $(N \times M) \times (G \times D)$ sparse matrix that maps the GRF values from the aggregated spatio--temporal dataset to the spatio--temporal mesh. The matrix $A$ has two important properties: first, its entries are non--negative, $A_{kp}^{ij} \geq 0$, and second, the rows sum to one, $\sum_{kp} A^{ij}_{kp} = 1$.  

The projection matrix $A$ can be interpreted as the product of two components: a spatial proportion, $|R_{ik}|/|R_i|$, and a temporal proportion, $|T_{jp}|/|T_j|$. If the spatial triangulated mesh is uniformly spaced over the study region and the temporal mesh partitions each interval uniformly, then the projection matrix simplifies to  
$A_{kp}^{ij} = 1/(n_i \cdot n_j)$,
where $n_i$ is the number of spatial mesh points lying inside region $R_i$, and $n_j$ is the number of temporal subdivisions of interval $T_j$. This simplification is exact for the temporal part and serves as a good approximation for the spatial part under uniform triangulations.

\section{Simulation study}
\label{sec:sim}

In this section, we conduct a simulation study to evaluate the performance of the proposed model. The synthetic data are generated from a spatio--temporal Gaussian process governed by a diffusion-based stochastic partial differential equation (SPDE), which can result in both separable and non--separable structures \cite{lindgren2023inlaspacetime}. The full formulation of the SPDE is provided in Section~\ref{sec:diffusion}. To assess model performance under varying levels of spatial and temporal aggregation, we consider 15 distinct scenarios. For each of these scenarios, we evaluate the model under three levels of temporal correlation: weak, moderate, and strong.
To contextualize these results, we benchmark our spatio--temporal disaggregation under a separable specification against a discrete areal baseline with a separable Besag--AR(1) random effect as defined in Section \ref{sec:areal_model}.

Model performance is measured by comparing the posterior mean estimates with the true simulated values using the Root Mean Squared Error (RMSE). We also evaluate predictive uncertainty by presenting the spatio--temporal empirical coverage probability together with credible interval width across all scenarios. Finally, we compute the empirical coverage of the 95$\%$ credible intervals for the simulation parameters.

\subsection{Baseline spatio--temporal areal model}
\label{sec:areal_model}

Let $\{y_{ij}\}$ denote the set of observations in regions $i=1,\ldots,N$ and time points $j=1,\ldots,M$. As a baseline to compare our spatio--temporal disaggregation model, we use an areal Gaussian model:
\begin{align}
    \begin{split}
        y_{ij} \mid \eta_{ij} &\sim \mathcal{N}(\eta_{ij},\,\tau_{\epsilon}^{-1}), \\
        \eta_{ij} &= \mathbf{x}_{ij}^\top \boldsymbol{\beta} \;+\; z_{ij},
    \end{split}
    \label{equ:areal-1}
\end{align}

Here, $\mathbf{x}_{ij}^\top \boldsymbol{\beta}$ represents the fixed effects including the intercept, $z_{ij}$ is the spatio--temporal structured random effect, and $\tau_{\epsilon}$ is the spatio--temporal precision for the unstructured random effect.
We assume that $z_{ij}$ has a separable covariance structure, 
which is a combination of a spatial Besag (ICAR) component, and a
temporal autoregressive process of order one, AR(1), following the framework of \cite{KnorrHeld2000,Wikle1998}.

Specifically, let $\mathbf{z}_j = (z_{1j},\ldots,z_{Nj})^\top$ denote the spatial random field at time $j$, and let $Q_S$ denote the Besag spatial precision matrix. The temporal evolution is then given by
\begin{align*}
    \begin{split}
        \mathbf{z}_j \mid \mathbf{z}_{j-1} &\sim 
        \mathcal{N}\!\big(\rho\,\mathbf{z}_{j-1},\; \tau_{S}^{-1} Q_S^{-}\big), 
        \quad j=2,\ldots,M,\\
        \mathbf{z}_1 &\sim 
        \mathcal{N}\!\big(\mathbf{0},\; \tau_{S}^{-1}(1-\rho^2)^{-1} Q_S^{-}\big),
    \end{split}
\end{align*}
where $|\rho|<1$ is the temporal autocorrelation parameter, $\tau_S$ is the spatial precision, and $Q_S^{-}$ denotes the Moore--Penrose pseudoinverse of $Q_S$. This specification implies a separable covariance structure of the form
$\text{Cov}(z_{ij}, z_{kl}) 
= \tau_S^{-1}\,\rho^{|j-l|}\, Q_S^{-}(i,k)$,
where $Q_S^{-}(i,k)$ denotes the $(i,k)$-th element of the pseudoinverse of the Besag precision matrix \cite{BlangiardoCameletti2015}.

\subsection{Simulation of continuous spatio--temporal processes}

The underlying continuous spatio--temporal process is defined over the spatial domain $\mathbf{s} \in [0,1]^2$ and the temporal interval $t \in [0, 24]$. For the simulation, the spatial domain is discretized into a regular $24 \times 24$ grid, yielding 576 uniformly spaced locations, while the temporal domain is divided into 24 equally spaced time points, denoted by $t \in {1, 2, \ldots, 24}$. Based on this discretization, we generate the following spatio--temporal continuous process:

\begin{equation}
    W(\mathbf{s},t) = \beta_0 + Z(\mathbf{s},t),
    \label{equ:cont-eq}
\end{equation}
where $\beta_0$ denotes the intercept and $Z(\mathbf{s})$ represents the spatio--temporal random effect simulated using the SPDE-based model described in Section~\ref{sec:diffusion}.
To ensure consistency of our results, we generated 50 independent realizations of the continuous spatio--temporal process.

The simulation was carried out with the \texttt{INLAspacetime} package \cite{lindgren2023inlaspacetime}, and the parameter values employed are summarized in Table~\ref{tab:param}.

\begin{table}[h!]
\centering
\begin{tabular}{cc}
\begin{minipage}{0.45\textwidth}
\centering
\begin{tabular}{|c|c|}
\hline
Parameter & Value \\ \hline
Intercept ($\beta_0$) & $0.1$ \\ \hline
Variance ($\sigma^2$) & $0.25$ \\ \hline
Spatial range ($r_s$) & $0.2$ \\ \hline
Standard error ($e$) & $0.15$ \\ \hline
\end{tabular}
\subcaption{Main simulation parameters.}
\label{tab:param_a}
\end{minipage}
&
\begin{minipage}{0.45\textwidth}
\centering
\begin{tabular}{|c|c|c|}
\hline
 & Separable & Non Separable \\ \hline
Weak   & $1$  & $2$  \\ \hline
Moderate & $3$  & $6$  \\ \hline
Strong & $12$ & $24$ \\ \hline
\end{tabular}
\subcaption{Temporal range ($r_t$).}
\label{tab:param_b}
\end{minipage}
\end{tabular}
\caption{Simulation setup used for generating the spatio--temporal random effect. (a) Core model parameters. (b) Temporal range values for weak, moderate, and strong correlations under both non--separable and separable settings.}
\label{tab:param}
\end{table}

We considered both separable and non--separable models to ensure that our approach to spatio--temporal disaggregation is robust and effective across different dependency structures.
Different values for temporal ranges were considered to emulate three levels of autocorrelation, namely, weak, moderate, and strong. This choice is motivated by the separable setting; we consider three temporal range parameters $\phi \in \{1,3,12\}$. Under the separable setting, the temporal correlation is exponential, $\mathrm{Corr}(h)=\exp(-h/\phi)$; sampling at unit steps ($\Delta=1$) induces an AR(1) process with
\begin{equation*}
\rho = \exp(-\Delta/\phi) = \exp(-1/\phi).
\end{equation*}
Thus, $\phi=\{1,3,12\}$ map to $\rho \approx \{0.37,\,0.72,\,0.92\}$, corresponding to weak, moderate, and strong autocorrelation, respectively.

\begin{figure}[h!]
\small
\centering
\begin{subfigure}{\textwidth}
    \centering
    \includegraphics[width=0.8\linewidth]{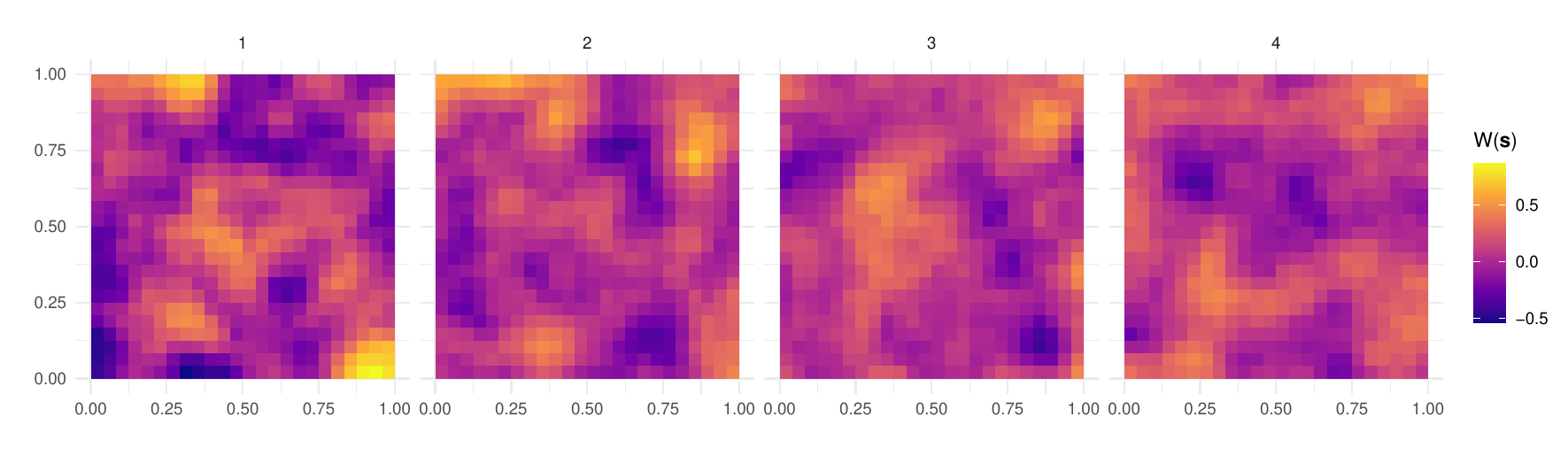}
    \vspace{-1em}
    \subcaption{Weak temporal autocorrelation}
    \vspace{-0.6em}
\end{subfigure}

\begin{subfigure}{\textwidth}
    \centering
    \includegraphics[width=0.8\linewidth]{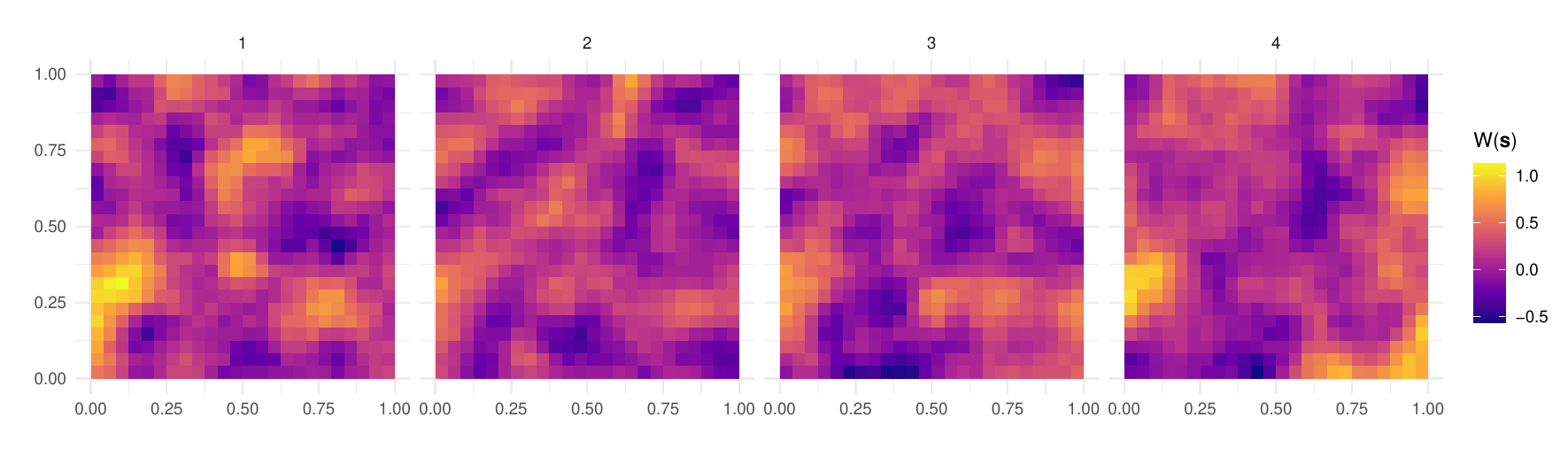}
    \vspace{-1em}
    \subcaption{Moderate temporal autocorrelation}
    \vspace{-0.8em}
\end{subfigure}

\begin{subfigure}{\textwidth}
    \centering
    \includegraphics[width=0.8\linewidth]{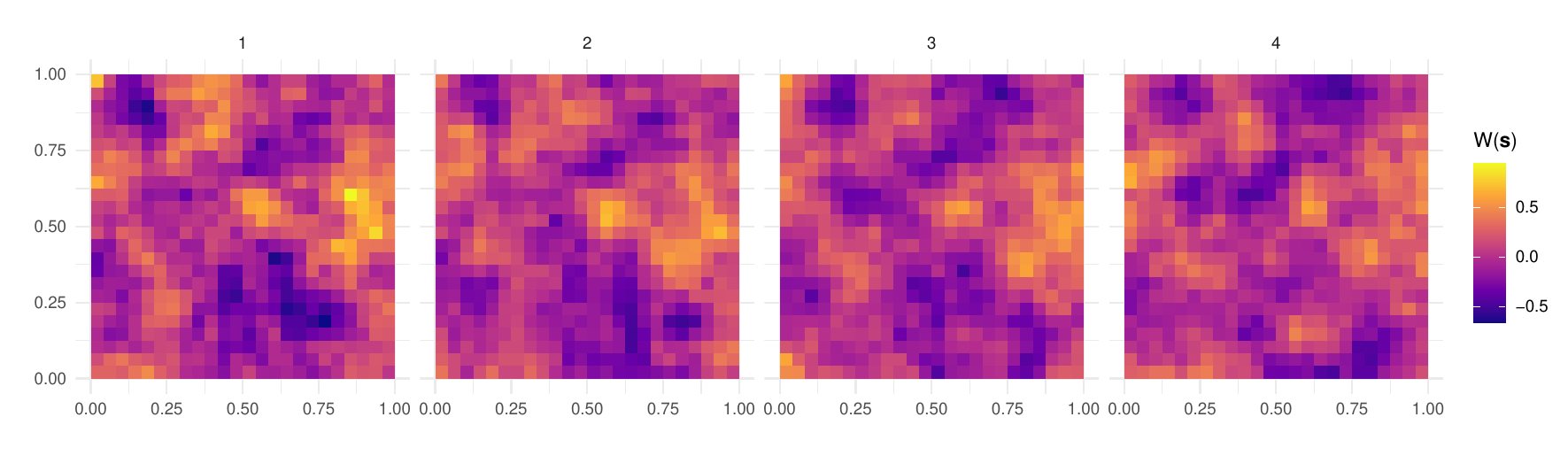}
    \vspace{-1em}
    \subcaption{Strong temporal autocorrelation}
\end{subfigure}

\caption{Realizations of the non--separable spatio--temporal process at the first four time points for the three temporal autocorrelation levels. }
\label{fig:sim-ex1}
\end{figure}

To maintain comparable levels of temporal dependence across model structures, we adjusted the temporal range values accordingly. As shown in Table~\ref{tab:param_b}, the values of the temporal range for the non--separable model are twice those of the separable one (see \cite{lindgren2023inlaspacetime}). This happens because the covariance structure for the non--separable model decays almost twice as fast as the separable one. Figure~\ref{fig:sim-ex1} illustrates one realization of the non--separable model, showing the evolution of the process across the first four time points for the three different temporal autocorrelation levels. 

\subsection{Spatio-temporal aggregated data scenarios}

To evaluate our model's performance under varying levels of coarsening, we constructed 15 distinct scenarios by combining different spatial and temporal aggregation settings. The spatial domain, initially represented by a $24 \times 24$ grid, was aggregated into coarser regular grids of size $3 \times 3$, $4 \times 4$, $6 \times 6$, $8 \times 8$, and $12 \times 12$, corresponding to reductions in resolution by compression factors of 8, 6, 4, 3, and 2, respectively. Similarly, the temporal domain, initially comprising 24 time points, was aggregated into 12, 8, or 6 intervals, corresponding to compression factors of 2, 3, and 4. Each scenario results from a unique combination of one spatial and one temporal aggregation level. For instance, under the finest aggregation configuration ($12 \times 12$ spatial grid and 12 temporal intervals), each aggregated value corresponds to the average of 4 spatial locations (a $2 \times 2$ block) and 2 consecutive time points, thus summarizing 8 original values into one. Table~\ref{tab:agg-summary} provides a complete overview of the number of original values averaged within each scenario.

\begin{table}[h!]
\centering
\small
\begin{tabular}{|c|c|c|c|c|c|}
\hline
\multirow{3}{*}{\begin{tabular}[c]{@{}c@{}}\textbf{Temporal} \\ \textbf{aggregation} \end{tabular}} & \multicolumn{5}{c|}{\textbf{Spatial resolution}} \\
\cline{2-6}
 & $3 \times 3$ & $4 \times 4$ & $6 \times 6$ & $8 \times 8$ & $12 \times 12$ \\
 & (factor 8) & (factor 6) & (factor 4) & (factor 3) & (factor 2) \\
\hline
12 (factor 2) & $8^2 \times 2 = 128$  & $6^2 \times 2 = 72$  & $4^2 \times 2 = 32$  & $3^2 \times 2 = 18$  & $2^2 \times 2 = 8$ \\
\hline
8  (factor 3) & $8^2 \times 3 = 192$  & $6^2 \times 3 = 108$ & $4^2 \times 3 = 48$  & $3^2 \times 3 = 27$  & $2^2 \times 3 = 12$ \\
\hline
6  (factor 4) & $8^2 \times 4 = 256$  & $6^2 \times 4 = 144$ & $4^2 \times 4 = 64$  & $3^2 \times 4 = 36$  & $2^2 \times 4 = 16$ \\
\hline
\end{tabular}
\caption{Number of observations aggregated in each spatio--temporal scenario. Each entry corresponds to the number of aggregated spatio--temporal points from the original simulated data in each scenario.}
\label{tab:agg-summary}
\end{table}

Once the 15 spatio--temporal aggregation scenarios were defined, with their corresponding number of spatial regions and temporal intervals, we proceeded to aggregate the continuous Gaussian processes from Equation~\ref{equ:cont-eq}. This was done by averaging the process values within each spatial region and over each temporal interval. To incorporate observational noise and simulate more realistic data, we then added independent and identically distributed (i.i.d.) zero-mean Gaussian noise to each spatio--temporal unit:

\begin{equation}
W(R_{ij}) = \beta_0 + \overline{Z}(R_{ij}) + e_{ij},
\label{equ:agg-1}
\end{equation}
where $e_{ij}$ denotes the i.i.d. Gaussian error term with zero mean and fixed standard deviation, as specified in Table~\ref{tab:param}. The term $\overline{Z}(R_{ij})$ represents the mean value of the latent process $Z(\mathbf{s}_k, t_m)$ over the spatio--temporal region $R_{ij}$ and is defined as

\begin{equation*}
\overline{Z}(R_{ij}) = \left( \frac{1}{|R_i||T_j|} \right) \sum_{\mathbf{s}_k \in R_i} \sum_{t_m \in T_j} Z(\mathbf{s}_k, t_m),
\end{equation*}
where $|R_i|$ and $|T_j|$ denote the number of spatial locations and time points included in region $R_i$ and interval $T_j$, respectively.

\begin{figure}[h!]
  \centering

  \begin{subfigure}[c]{0.33\textwidth}
    \includegraphics[width=0.9\linewidth]{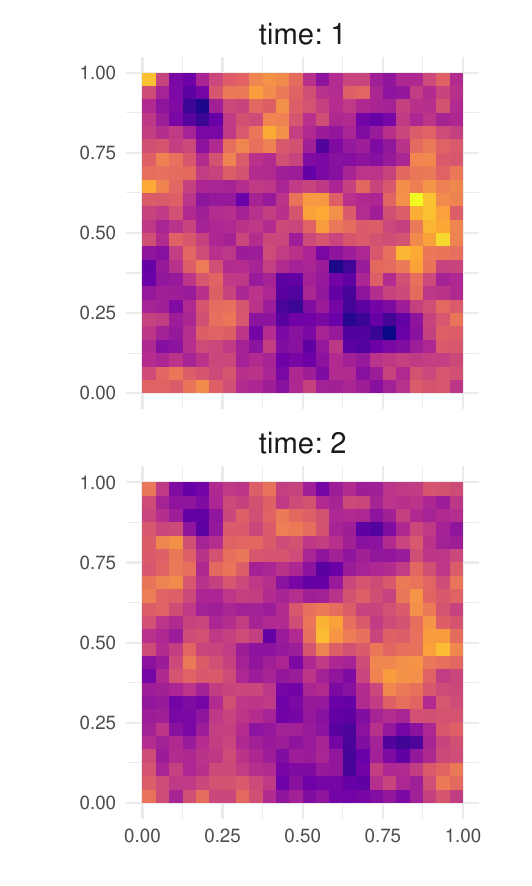}
    \caption{Continuous process $W(\mathbf{s},t)$}
    \label{fig:ex2-cont}
  \end{subfigure}
  \hfill
  \begin{subfigure}[c]{0.65\textwidth}
    \centering
    \begin{minipage}[t]{\linewidth}
      \begin{minipage}[t]{0.32\linewidth}
        \includegraphics[width=\linewidth]{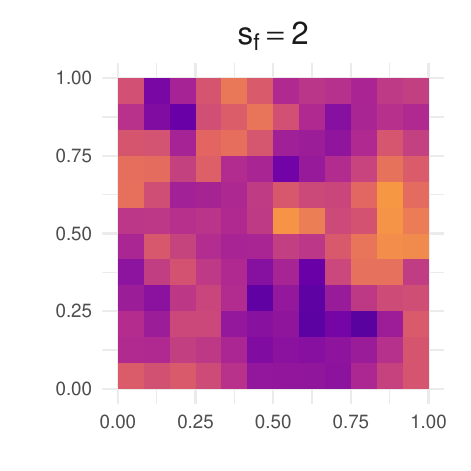}
      \end{minipage}\hfill
      \begin{minipage}[t]{0.32\linewidth}
        \includegraphics[width=\linewidth]{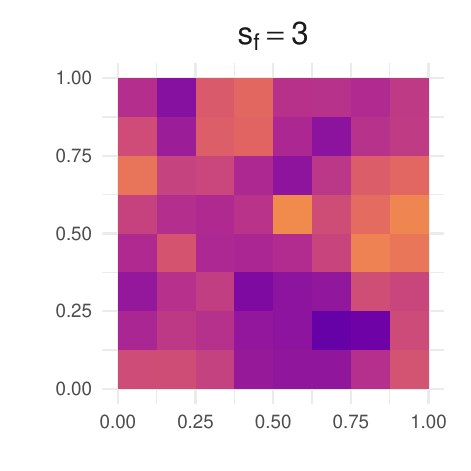}
      \end{minipage}\hfill
      \begin{minipage}[t]{0.32\linewidth}
        \includegraphics[width=\linewidth]{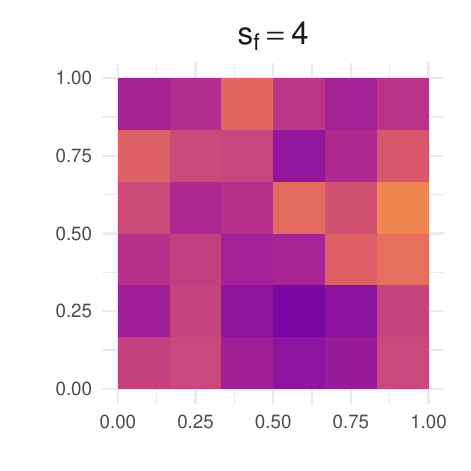}
      \end{minipage}
    \end{minipage}
    \vspace{0.5em}

    \begin{minipage}[t]{\linewidth}
    \begin{minipage}[t]{0.15\linewidth}
        \color{white}{hello}
      \end{minipage}
      \hfill
      \begin{minipage}[t]{0.32\linewidth}
        \includegraphics[width=\linewidth]{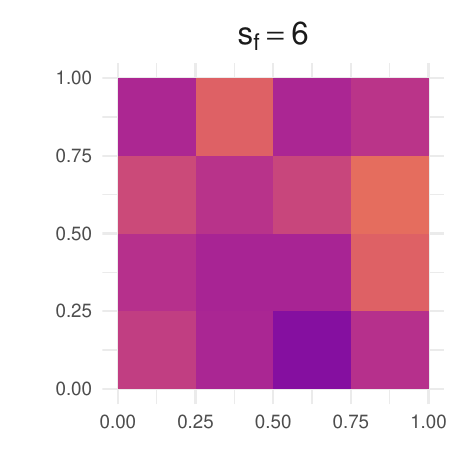}
      \end{minipage}
      \hfill
      \begin{minipage}[t]{0.32\linewidth}
        \includegraphics[width=\linewidth]{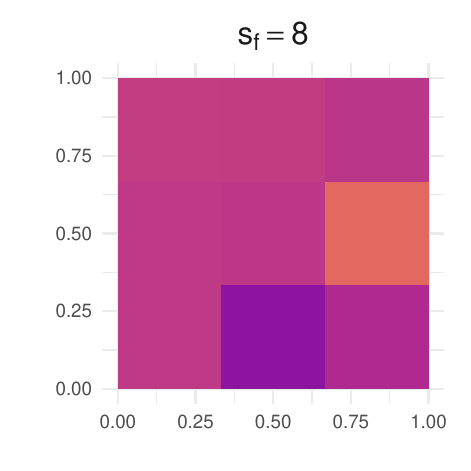}
      \end{minipage}
      \hfill
      \begin{minipage}[t]{0.15\linewidth}
        \color{white}{hello}
      \end{minipage}
    \end{minipage}
    \caption{\small Aggregated process $W(R_{ij})$}
    \label{fig:ex2-agg}
  \end{subfigure}
  \caption{(a) Realization of the continuous process $W(\mathbf{s})$ in the non--separable setting when the temporal autocorrelation is strong, and (b) its aggregated versions $W(R_{ij})$ obtained using different spatial configurations, with temporal compression aggregation fixed at 2.}
  \label{fig:ex2-main}
\end{figure}

To visualize the effect of spatio--temporal aggregation, Figure~\ref{fig:ex2-main} displays the five spatial grid configurations corresponding to the first row of Table~\ref{tab:agg-summary}, where the temporal resolution is fixed at 12 intervals for the non--separable setting. In this case, Figure~\ref{fig:ex2-cont} represents the original simulated data, and Figure~\ref{fig:ex2-agg} presents the aggregated data when the temporal compression factor is 2. 

\subsection{Simulation Results}

\subsubsection{Separable model}

We present the results for the separable simulated data and compare the performance of the spatio--temporal disaggregation model with that of a classical spatio--temporal areal model. To ensure a fair comparison, the areal model is also specified with the separable model that was defined in Section~\ref{sec:areal_model}. For disaggregation, the areal model assumes that values remain constant within each spatial unit (square) of the original grid and are duplicated across the corresponding time intervals. For instance, with spatial and temporal aggregation factors of 2, each aggregated value in the areal model is uniformly distributed across the 8 corresponding locations in the finer-resolution grid.

\begin{table}[h!]
\centering
\begin{tabular}{cc|c|c|c|c|c|c|}
\cline{3-8}
 &  & \multicolumn{2}{c|}{\begin{tabular}[c]{@{}c@{}}Weak\\ Autocorrelation\end{tabular}} & 
       \multicolumn{2}{c|}{\begin{tabular}[c]{@{}c@{}}Moderate\\ Autocorrelation\end{tabular}} & 
       \multicolumn{2}{c|}{\begin{tabular}[c]{@{}c@{}}Strong\\ Autocorrelation\end{tabular}} \\ \hline
\multicolumn{1}{|c|}{\begin{tabular}[c]{@{}c@{}}Spatial\\ Factor ($s_f$)\end{tabular}} & 
\begin{tabular}[c]{@{}c@{}}Temporal\\ Factor ($t_f$)\end{tabular} & 
\multicolumn{1}{c|}{\begin{tabular}[c]{@{}c@{}}Areal\\ Model\end{tabular}} & 
\multicolumn{1}{c|}{\begin{tabular}[c]{@{}c@{}}Cont\\ Model\end{tabular}} & 
\multicolumn{1}{c|}{\begin{tabular}[c]{@{}c@{}}Areal\\ Model\end{tabular}} & 
\multicolumn{1}{c|}{\begin{tabular}[c]{@{}c@{}}Cont\\ Model\end{tabular}} & 
\multicolumn{1}{c|}{\begin{tabular}[c]{@{}c@{}}Areal\\ Model\end{tabular}} & 
\multicolumn{1}{c|}{\begin{tabular}[c]{@{}c@{}}Cont\\ Model\end{tabular}} \\ \hline
\multicolumn{1}{|c|}{2} & 2 & 0.184 & 0.175 & 0.189 & 0.172 & 0.165 & 0.143 \\ \hline
\multicolumn{1}{|c|}{2} & 3 & 0.197 & 0.191 & 0.204 & 0.191 & 0.175 & 0.155 \\ \hline
\multicolumn{1}{|c|}{2} & 4 & 0.203 & 0.199 & 0.213 & 0.203 & 0.183 & 0.163 \\ \hline
\multicolumn{1}{|c|}{3} & 2 & 0.197 & 0.189 & 0.209 & 0.192 & 0.192 & 0.171 \\ \hline
\multicolumn{1}{|c|}{3} & 3 & 0.205 & 0.201 & 0.221 & 0.208 & 0.200 & 0.180 \\ \hline
\multicolumn{1}{|c|}{3} & 4 & 0.209 & 0.208 & 0.227 & 0.217 & 0.206 & 0.188 \\ \hline
\multicolumn{1}{|c|}{4} & 2 & 0.207 & 0.206 & 0.224 & 0.209 & 0.210 & 0.192 \\ \hline
\multicolumn{1}{|c|}{4} & 3 & 0.211 & 0.214 & 0.232 & 0.223 & 0.217 & 0.200 \\ \hline
\multicolumn{1}{|c|}{4} & 4 & 0.213 & 0.220 & 0.237 & 0.229 & 0.223 & 0.207 \\ \hline
\multicolumn{1}{|c|}{6} & 2 & 0.213 & 0.221 & 0.239 & 0.232 & 0.233 & 0.222 \\ \hline
\multicolumn{1}{|c|}{6} & 3 & 0.215 & 0.232 & 0.243 & 0.240 & 0.237 & 0.227 \\ \hline
\multicolumn{1}{|c|}{6} & 4 & 0.216 & 0.238 & 0.245 & 0.247 & 0.241 & 0.233 \\ \hline
\multicolumn{1}{|c|}{8} & 2 & 0.215 & 0.232 & 0.244 & 0.246 & 0.246 & 0.241 \\ \hline
\multicolumn{1}{|c|}{8} & 3 & 0.219 & 0.235 & 0.247 & 0.251 & 0.250 & 0.246 \\ \hline
\multicolumn{1}{|c|}{8} & 4 & 0.219 & 0.246 & 0.249 & 0.258 & 0.253 & 0.251 \\ \hline
\end{tabular}
\caption{Root mean square error (RMSE) between model predictions and true values, comparing the Areal model and the proposed continuous spatio--temporal disaggregation model in the separable scenario. The evaluation is conducted under weak, moderate, and strong temporal autocorrelation structures. Results are reported for 15 aggregated spatio--temporal scenarios, obtained from combinations of spatial ($s_f$) and temporal ($t_f$) aggregation factors.}
\label{tab:res-sep-rmse}
\end{table}

Table~\ref{tab:res-sep-rmse} shows the average RMSE  across different scenarios for both our spatio--temporal disaggregation model and the areal model. We observe that, for a fixed spatial aggregation factor, in our spatio--temporal continuous disaggregation model, the RMSE increases as the temporal aggregation factor increases. This indicates that the approximation becomes less accurate as the temporal resolution becomes coarser, which is expected since less information is available. Similarly, when the temporal factor is held constant and the spatial aggregation factor increases, we see the same behavior. These patterns confirm that the model's ability to recover the original data depends on the level of spatial and temporal aggregation: the finer the resolution, the better the approximation. 

As shown in Table~\ref{tab:res-sep-rmse}, the performance of the continuous disaggregation model relative to the areal baseline depends on the temporal autocorrelation structure. Under weak temporal autocorrelation, the continuous disaggregation model achieves lower RMSE for spatial aggregation factors 2 and 3 across all temporal factors. For moderate temporal autocorrelation, the improvement extends to spatial aggregation factors 2, 3, and 4, while under strong temporal autocorrelation, it includes all spatial and temporal aspects. These results indicate that the advantage of the continuous disaggregation model increases as temporal autocorrelation strengthens. Moreover, the lowest RMSE values are observed when temporal autocorrelation is strong and both spatial and temporal resolutions are high, suggesting that the proposed model is particularly suitable for applications characterized by strong temporal autocorrelation and low spatial and temporal aggregation factors.

We present the results of the parameter estimation used in the simulation. For each parameter, we computed two-sided $95\%$ credible intervals and calculated the proportion of the 50 simulations in which the true parameter value was contained within the interval. Table~\ref{tab:cov-sep} reports the average coverage rate for each scenario. Fixed parameters, such as the intercept ($\beta_0$), exhibit consistently high coverage across all scenarios and temporal autocorrelation structures. For Gaussian process hyperparameters, standard error ($e$), spatial range ($r_s$), temporal range ($r_t$), and variance ($\sigma^2$) coverage depends on the temporal autocorrelation: under weak autocorrelation, acceptable coverage occurs for $s_f \in \{2,3\}$; under moderate autocorrelation, for $s_f \in \{2,3,4\}$; and under strong autocorrelation, for all spatial and temporal aggregation factors. This suggests that parameter estimation for Gaussian processes depends on temporal autocorrelation; higher autocorrelation leads to better parameter estimation. 

\begin{table}[h]
\centering
\scriptsize
\resizebox{\textwidth}{!}{%
\begin{tabular}{|c|c|ccccc|ccccc|ccccc|}
\hline
\multirow{2}{*}{$s_f$} & \multirow{2}{*}{$t_f$} & \multicolumn{5}{c|}{Weak Autocorrelation} & \multicolumn{5}{c|}{Moderate Autocorrelation} & \multicolumn{5}{c|}{Strong Autocorrelation} \\ \cline{3-17}
 & & SE & $r_s$ & $r_t$ & Var & Int. & SE & $r_s$ & $r_t$ & Var & Int. & SE & $r_s$ & $r_t$ & Var & Int. \\ \hline
2 & 2 & \cellcolor{yellow!25}0.94 & \cellcolor{yellow!25}0.94 & \cellcolor{green!25}0.96 & \cellcolor{green!25}0.96 & \cellcolor{green!25}0.98 & \cellcolor{green!25}0.98 & \cellcolor{yellow!25}0.94 & \cellcolor{green!25}0.96 & \cellcolor{green!25}0.98 & \cellcolor{green!25}0.98 & \cellcolor{yellow!25}0.94 & \cellcolor{green!25}0.96 & \cellcolor{green!25}1.00 & \cellcolor{yellow!25}0.94 & \cellcolor{yellow!25}0.94 \\ \hline
2 & 3 & \cellcolor{green!25}0.98 & \cellcolor{yellow!25}0.94 & \cellcolor{green!25}0.96 & \cellcolor{green!25}0.98 & \cellcolor{green!25}0.98 & \cellcolor{yellow!25}0.94 & \cellcolor{yellow!25}0.94 & \cellcolor{green!25}0.96 & \cellcolor{green!25}0.98 & \cellcolor{green!25}1.00 & \cellcolor{yellow!25}0.94 & \cellcolor{green!25}0.96 & \cellcolor{green!25}0.96 & \cellcolor{green!25}0.96 & \cellcolor{yellow!25}0.90 \\ \hline
2 & 4 & \cellcolor{yellow!25}0.92 & \cellcolor{yellow!25}0.94 & \cellcolor{green!25}0.96 & \cellcolor{yellow!25}0.94 & \cellcolor{green!25}1.00 & \cellcolor{green!25}0.98 & \cellcolor{yellow!25}0.94 & \cellcolor{yellow!25}0.94 & \cellcolor{green!25}0.96 & \cellcolor{yellow!25}0.96 & \cellcolor{green!25}1.00 & \cellcolor{yellow!25}0.94 & \cellcolor{green!25}0.98 & \cellcolor{yellow!25}0.94 & \cellcolor{yellow!25}0.86 \\ \hline
3 & 2 & \cellcolor{green!25}0.98 & \cellcolor{yellow!25}0.92 & \cellcolor{green!25}0.98 & \cellcolor{green!25}0.96 & \cellcolor{green!25}0.96 & \cellcolor{green!25}0.98 & \cellcolor{yellow!25}0.94 & \cellcolor{yellow!25}0.94 & \cellcolor{green!25}0.98 & \cellcolor{yellow!25}0.96 & \cellcolor{green!25}0.98 & \cellcolor{yellow!25}0.94 & \cellcolor{yellow!25}0.88 & \cellcolor{green!25}0.96 & \cellcolor{yellow!25}0.94 \\ \hline
3 & 3 & \cellcolor{yellow!25}0.94 & \cellcolor{green!25}0.96 & \cellcolor{green!25}1.00 & \cellcolor{yellow!25}0.92 & \cellcolor{green!25}1.00 & \cellcolor{yellow!25}0.94 & \cellcolor{yellow!25}0.94 & \cellcolor{green!25}1.00 & \cellcolor{green!25}1.00 & \cellcolor{yellow!25}0.96 & \cellcolor{green!25}0.98 & \cellcolor{green!25}0.96 & \cellcolor{yellow!25}0.94 & \cellcolor{yellow!25}0.94 & \cellcolor{yellow!25}0.86 \\ \hline
3 & 4 & \cellcolor{yellow!25}0.92 & \cellcolor{yellow!25}0.90 & \cellcolor{green!25}0.96 & \cellcolor{yellow!25}0.92 & \cellcolor{green!25}0.98 & \cellcolor{green!25}1.00 & \cellcolor{yellow!25}0.90 & \cellcolor{green!25}1.00 & \cellcolor{green!25}1.00 & \cellcolor{yellow!25}0.96 & \cellcolor{green!25}1.00 & \cellcolor{yellow!25}0.92 & \cellcolor{green!25}1.00 & \cellcolor{green!25}1.00 & \cellcolor{yellow!25}0.88 \\ \hline
4 & 2 & \cellcolor{yellow!25}0.80 & \cellcolor{red!25}0.72 & \cellcolor{green!25}1.00 & \cellcolor{red!25}0.70 & \cellcolor{green!25}0.96 & \cellcolor{green!25}1.00 & \cellcolor{green!25}1.00 & \cellcolor{green!25}0.98 & \cellcolor{green!25}0.98 & \cellcolor{yellow!25}0.96 & \cellcolor{yellow!25}0.92 & \cellcolor{green!25}0.98 & \cellcolor{yellow!25}0.94 & \cellcolor{green!25}0.96 & \cellcolor{yellow!25}0.86 \\ \hline
4 & 3 & \cellcolor{red!25}0.70 & \cellcolor{red!25}0.70 & \cellcolor{green!25}0.96 & \cellcolor{yellow!25}0.72 & \cellcolor{green!25}1.00 & \cellcolor{yellow!25}0.86 & \cellcolor{yellow!25}0.86 & \cellcolor{yellow!25}0.94 & \cellcolor{yellow!25}0.86 & \cellcolor{yellow!25}0.94 & \cellcolor{yellow!25}0.94 & \cellcolor{green!25}0.98 & \cellcolor{green!25}1.00 & \cellcolor{yellow!25}0.94 & \cellcolor{yellow!25}0.92 \\ \hline
4 & 4 & \cellcolor{red!25}0.74 & \cellcolor{red!25}0.68 & \cellcolor{yellow!25}0.88 & \cellcolor{red!25}0.66 & \cellcolor{green!25}0.96 & \cellcolor{yellow!25}0.94 & \cellcolor{green!25}0.96 & \cellcolor{green!25}0.98 & \cellcolor{yellow!25}0.92 & \cellcolor{yellow!25}0.96 & \cellcolor{green!25}0.98 & \cellcolor{green!25}0.96 & \cellcolor{green!25}0.96 & \cellcolor{green!25}1.00 & \cellcolor{yellow!25}0.90 \\ \hline
6 & 2 & \cellcolor{red!25}0.50 & \cellcolor{red!25}0.66 & \cellcolor{yellow!25}0.92 & \cellcolor{red!25}0.54 & \cellcolor{green!25}0.98 & \cellcolor{red!25}0.78 & \cellcolor{green!25}0.96 & \cellcolor{yellow!25}0.90 & \cellcolor{yellow!25}0.90 & \cellcolor{yellow!25}0.94 & \cellcolor{yellow!25}0.94 & \cellcolor{green!25}0.98 & \cellcolor{green!25}0.98 & \cellcolor{green!25}0.98 & \cellcolor{yellow!25}0.88 \\ \hline
6 & 3 & \cellcolor{red!25}0.54 & \cellcolor{red!25}0.66 & \cellcolor{green!25}0.96 & \cellcolor{red!25}0.46 & \cellcolor{green!25}0.98 & \cellcolor{yellow!25}0.82 & \cellcolor{yellow!25}0.88 & \cellcolor{green!25}0.96 & \cellcolor{yellow!25}0.86 & \cellcolor{green!25}0.98 & \cellcolor{yellow!25}0.94 & \cellcolor{green!25}0.98 & \cellcolor{yellow!25}0.92 & \cellcolor{green!25}0.98 & \cellcolor{yellow!25}0.90 \\ \hline
6 & 4 & \cellcolor{red!25}0.44 & \cellcolor{yellow!25}0.84 & \cellcolor{yellow!25}0.90 & \cellcolor{red!25}0.44 & \cellcolor{green!25}0.98 & \cellcolor{red!25}0.68 & \cellcolor{green!25}0.98 & \cellcolor{green!25}0.96 & \cellcolor{red!25}0.78 & \cellcolor{green!25}1.00 & \cellcolor{yellow!25}0.92 & \cellcolor{green!25}0.96 & \cellcolor{yellow!25}0.90 & \cellcolor{yellow!25}0.92 & \cellcolor{yellow!25}0.94 \\ \hline
8 & 2 & \cellcolor{red!25}0.54 & \cellcolor{green!25}1.00 & \cellcolor{green!25}0.96 & \cellcolor{red!25}0.52 & \cellcolor{green!25}0.98 & \cellcolor{red!25}0.76 & \cellcolor{green!25}1.00 & \cellcolor{yellow!25}0.86 & \cellcolor{yellow!25}0.80 & \cellcolor{green!25}1.00 & \cellcolor{yellow!25}0.90 & \cellcolor{green!25}1.00 & \cellcolor{green!25}0.96 & \cellcolor{green!25}0.96 & \cellcolor{yellow!25}0.92 \\ \hline
8 & 3 & \cellcolor{red!25}0.52 & \cellcolor{green!25}0.98 & \cellcolor{yellow!25}0.92 & \cellcolor{red!25}0.52 & \cellcolor{green!25}0.98 & \cellcolor{red!25}0.70 & \cellcolor{green!25}0.98 & \cellcolor{yellow!25}0.84 & \cellcolor{red!25}0.78 & \cellcolor{yellow!25}0.96 & \cellcolor{yellow!25}0.94 & \cellcolor{green!25}0.98 & \cellcolor{yellow!25}0.92 & \cellcolor{yellow!25}0.96 & \cellcolor{yellow!25}0.88 \\ \hline
8 & 4 & \cellcolor{red!25}0.32 & \cellcolor{green!25}1.00 & \cellcolor{yellow!25}0.90 & \cellcolor{red!25}0.38 & \cellcolor{green!25}1.00 & \cellcolor{red!25}0.64 & \cellcolor{green!25}0.98 & \cellcolor{green!25}1.00 & \cellcolor{red!25}0.68 & \cellcolor{green!25}1.00 & \cellcolor{yellow!25}0.88 & \cellcolor{green!25}1.00 & \cellcolor{yellow!25}0.94 & \cellcolor{yellow!25}0.94 & \cellcolor{yellow!25}0.86 \\ \hline
\end{tabular}%
}
\caption{Coverage of $95\ \%$ parameters credible intervals across all scenarios. Here, $s_f$ = spatial factor, $t_f$ = temporal factor, SE = standard error ($e$), $r_s$ = spatial range, $r_t$ = temporal range, Var = variance ($\sigma^2$), and Int. = intercept ($\beta_0$). Green denotes coverage $\geq 95\ \%$, yellow denotes $80\ \% \leq \text{coverage} < 95\ \%$, and red denotes $\text{coverage} < 80\ \%$.}
\label{tab:cov-sep}
\end{table}

Spatio--temporal coverages are reported in Table~\ref{tab:coverage1}. Each entry is the empirical coverage probability (ECP) of the nominal two-sided $95\%$ posterior credible interval, that is, the proportion of space--time cells in which the credible interval contains the true value. Under weak and moderate temporal autocorrelation, ECP generally increases with the spatial factor $s_f$, attaining its highest levels at $s_f=8$. Under strong temporal autocorrelation, ECP is overall stable across all scenarios. However, because high coverage can be achieved simply by widening intervals, coverage alone is not diagnostic.

\begin{table}[h!]
\small
\centering
\setlength{\tabcolsep}{4pt}
\renewcommand{\arraystretch}{1.1}
\begin{tabular}{l *{5}{c} *{5}{c} *{5}{c}}
\toprule
& \multicolumn{5}{c}{Weak Autocorrelation} & \multicolumn{5}{c}{Moderate Autocorrelation} & \multicolumn{5}{c}{Strong Autocorrelation} \\
\cmidrule(lr){2-6}\cmidrule(lr){7-11}\cmidrule(lr){12-16}
$t_f \setminus s_f$ & 2 & 3 & 4 & 6 & 8 & 2 & 3 & 4 & 6 & 8 & 2 & 3 & 4 & 6 & 8 \\
\midrule
2 & 0.94 & 0.93 & 0.94 & 0.85 & 0.96 & 0.94 & 0.94 & 0.93 & 0.90 & 0.95 & 0.93 & 0.93 & 0.94 & 0.91 & 0.93 \\
3 & 0.93 & 0.92 & 0.92 & 0.92 & 0.96 & 0.93 & 0.93 & 0.92 & 0.91 & 0.96 & 0.93 & 0.94 & 0.94 & 0.96 & 0.94 \\
4 & 0.92 & 0.93 & 0.90 & 0.97 & 0.99 & 0.94 & 0.93 & 0.92 & 0.93 & 0.97 & 0.93 & 0.94 & 0.93 & 0.93 & 0.93 \\
\bottomrule
\end{tabular}
\caption{Empirical coverage probability (ECP) of nominal two-sided $95\%$ posterior credible intervals. Rows index the temporal factor $t_f$; columns the spatial factor $s_f$ for the separable model.}
\label{tab:coverage1}
\end{table}

To complement this, Table~\ref{tab:width1} reports the mean width of the $95\%$ credible intervals. As expected, interval width increases with finer spatial or temporal resolution with means larger $s_f$ or $t_f$, and the smallest widths occur at the coarsest setting $(s_f,t_f)=(2,2)$ for all temporal autocorrelation.

\begin{table}[h!]
\small
\centering
\setlength{\tabcolsep}{4pt}
\renewcommand{\arraystretch}{1.1}
\begin{tabular}{l *{5}{c} *{5}{c} *{5}{c}}
\toprule
& \multicolumn{5}{c}{Weak Autocorrelation} & \multicolumn{5}{c}{Moderate Autocorrelation} & \multicolumn{5}{c}{Strong Autocorrelation} \\
\cmidrule(lr){2-6}\cmidrule(lr){7-11}\cmidrule(lr){12-16}
$t_f \setminus s_f$ & 2 & 3 & 4 & 6 & 8 & 2 & 3 & 4 & 6 & 8 & 2 & 3 & 4 & 6 & 8 \\
\midrule
2 & 0.65 & 0.72 & 0.96 & 1.04 & 1.32 & 0.64 & 0.74 & 0.86 & 1.18 & 1.38 & 0.51 & 0.63 & 0.75 & 0.91 & 1.05 \\
3 & 0.70 & 0.77 & 1.05 & 1.34 & 1.39 & 0.71 & 0.81 & 1.04 & 1.22 & 1.40 & 0.57 & 0.68 & 0.81 & 1.09 & 1.29 \\
4 & 0.73 & 0.85 & 1.07 & 1.49 & 1.57 & 0.78 & 0.87 & 1.09 & 1.30 & 1.54 & 0.60 & 0.74 & 0.88 & 1.22 & 1.19 \\
\bottomrule
\end{tabular}
\caption{Mean width of the two-sided $95\%$ posterior credible intervals, averaged over space and time. Rows index $t_f$, columns index $s_f$ within each temporal-autocorrelation configuration for the separable model.}
\label{tab:width1}
\end{table}

Taken together, the two tables indicate a trade-off under weak or moderate temporal autocorrelation; the higher coverage at $s_f=8$ coincides with wider intervals, reflecting limited temporal borrowing and a tendency of our disaggregation model to average within regions. Under strong temporal autocorrelation, intervals are narrower at comparable $(s_f,t_f)$, suggesting more effective model capabilities to borrow information and, with that, better recovery of the continuous latent field.

Finally, in Appendix \ref{sec:appendix-sim-res} we present some results for the separable model in Figures \ref{fig:appendix-sep-1}, \ref{fig:appendix-sep-2} and \ref{fig:appendix-sep-3} when the spatial and temporal aggregation factors are 2.

\subsubsection{Non separable model}

Here we present the simulation results for the non--separable case using our spatio--temporal disaggregation model. Unlike the separable case, we do not compare with other models because, to the best of our knowledge, there are not readily available software to fit non--separable spatio--temporal areal models.
Instead, we compare our model’s predictions directly with the true values.
In Appendix~\ref{sec:appendix-sim-res}, we present results for the non--separable model (Figures~\ref{fig:appendix-non-sep-1}-\ref{fig:appendix-non-sep-3}) for the case where both the spatial and temporal aggregation factors are set to $2$.

Table~\ref{tab:res-nonsep-rmse} shows the average RMSE across different scenarios. The results reveal the same pattern observed in the separable model: for a fixed spatial aggregation factor, RMSE increases as the temporal aggregation factor grows, and vice versa. This confirms that, in both cases, the model's ability to recover the original data declines as aggregation increases, with finer spatial and temporal resolutions leading to more accurate approximations.

\begin{table}[h!]
\centering
\resizebox{\textwidth}{!}{%
\begin{tabular}{|c|c|c|c|c|}
\hline
\begin{tabular}[c]{@{}c@{}}Spatial\\ Factor ($s_f$)\end{tabular} &
\begin{tabular}[c]{@{}c@{}}Temporal\\ Factor ($t_f$)\end{tabular} &
\begin{tabular}[c]{@{}c@{}}Weak\\ Autocorrelation\end{tabular} &
\begin{tabular}[c]{@{}c@{}}Moderate\\ Autocorrelation\end{tabular} &
\begin{tabular}[c]{@{}c@{}}Strong\\ Autocorrelation\end{tabular} \\ \hline
2 & 2 & 0.1526 & 0.1544 & 0.1406 \\ \hline
2 & 3 & 0.1713 & 0.1711 & 0.1511 \\ \hline
2 & 4 & 0.1810 & 0.1834 & 0.1591 \\ \hline
3 & 2 & 0.1667 & 0.1732 & 0.1641 \\ \hline
3 & 3 & 0.1808 & 0.1877 & 0.1732 \\ \hline
3 & 4 & 0.1896 & 0.1975 & 0.1803 \\ \hline
4 & 2 & 0.1799 & 0.1895 & 0.1845 \\ \hline
4 & 3 & 0.1946 & 0.2025 & 0.1926 \\ \hline
4 & 4 & 0.1982 & 0.2104 & 0.1981 \\ \hline
6 & 2 & 0.2041 & 0.2129 & 0.2130 \\ \hline
6 & 3 & 0.2104 & 0.2238 & 0.2196 \\ \hline
6 & 4 & 0.2199 & 0.2317 & 0.2241 \\ \hline
8 & 2 & 0.2158 & 0.2317 & 0.2304 \\ \hline
8 & 3 & 0.2228 & 0.2355 & 0.2346 \\ \hline
8 & 4 & 0.2255 & 0.2429 & 0.2405 \\ \hline
\end{tabular}%
}
\caption{Root mean square error (RMSE) between predictions from the continuous spatio--temporal disaggregation model and the true values in the non--separable scenario. Results are shown for weak, moderate, and strong temporal autocorrelation structures across 15 aggregated spatio--temporal scenarios, obtained from combinations of spatial ($s_f$) and temporal ($t_f$) aggregation factors.}
\label{tab:res-nonsep-rmse}
\end{table}

For the non--separable case (Table~\ref{tab:coverage-nonsep}), we report the $95\%$ coverage rates, calculated as the proportion of 50 simulations in which the true parameter value was contained within the credible interval, across 15 scenarios. As observed for the separable structure, the intercept ($\beta_0$) maintains consistently high coverage across all scenarios and temporal autocorrelation levels. For the Gaussian process hyperparameters—standard error ($e$), spatial range ($r_s$), temporal range ($r_t$), and variance ($\sigma^2$) coverage patterns again depend on the temporal autocorrelation: under weak autocorrelation, acceptable coverage occurs for $s_f \in \{2,3,4\}$; under moderate autocorrelation, for $s_f \in \{2,3,4\}$; and under strong autocorrelation, for all spatial and temporal aggregation factors. These results reinforce that temporal autocorrelation plays a key role in accurate Gaussian process parameter recovery, regardless of whether the model presents a separable or non--separable structure. 

\begin{table}[h!]
\centering
\scriptsize
\resizebox{\textwidth}{!}{%
\begin{tabular}{|c|c|ccccc|ccccc|ccccc|}
\hline
\multirow{2}{*}{$s_f$} & \multirow{2}{*}{$t_f$} & \multicolumn{5}{c|}{Weak Autocorrelation} & \multicolumn{5}{c|}{Moderate Autocorrelation} & \multicolumn{5}{c|}{Strong Autocorrelation} \\ \cline{3-17}
 &  & SE & $r_s$ & $r_t$ & Var & Int. & SE & $r_s$ & $r_t$ & Var & Int. & SE & $r_s$ & $r_t$ & Var & Int. \\ \hline
2 & 2 & \cellcolor{green!25}1.00 & \cellcolor{green!25}0.98 & \cellcolor{green!25}1.00 & \cellcolor{green!25}0.98 & \cellcolor{yellow!25}0.92 & \cellcolor{green!25}1.00 & \cellcolor{green!25}1.00 & \cellcolor{green!25}0.96 & \cellcolor{green!25}0.98 & \cellcolor{yellow!25}0.90 & \cellcolor{green!25}0.98 & \cellcolor{green!25}0.98 & \cellcolor{green!25}0.96 & \cellcolor{yellow!25}0.94 & \cellcolor{yellow!25}0.90 \\ \hline
2 & 3 & \cellcolor{yellow!25}0.92 & \cellcolor{yellow!25}0.92 & \cellcolor{green!25}0.98 & \cellcolor{yellow!25}0.94 & \cellcolor{yellow!25}0.90 & \cellcolor{green!25}0.98 & \cellcolor{green!25}0.96 & \cellcolor{yellow!25}0.94 & \cellcolor{yellow!25}0.94 & \cellcolor{yellow!25}0.88 & \cellcolor{green!25}0.98 & \cellcolor{yellow!25}0.94 & \cellcolor{green!25}1.00 & \cellcolor{green!25}0.98 & \cellcolor{yellow!25}0.90 \\ \hline
2 & 4 & \cellcolor{yellow!25}0.88 & \cellcolor{green!25}0.96 & \cellcolor{green!25}1.00 & \cellcolor{green!25}1.00 & \cellcolor{yellow!25}0.92 & \cellcolor{green!25}0.98 & \cellcolor{green!25}0.96 & \cellcolor{green!25}1.00 & \cellcolor{green!25}1.00 & \cellcolor{yellow!25}0.92 & \cellcolor{green!25}1.00 & \cellcolor{yellow!25}0.92 & \cellcolor{green!25}1.00 & \cellcolor{green!25}0.98 & \cellcolor{yellow!25}0.92 \\ \hline
3 & 2 & \cellcolor{green!25}0.96 & \cellcolor{yellow!25}0.90 & \cellcolor{green!25}0.98 & \cellcolor{yellow!25}0.92 & \cellcolor{yellow!25}0.92 & \cellcolor{yellow!25}0.92 & \cellcolor{green!25}0.98 & \cellcolor{green!25}0.96 & \cellcolor{green!25}1.00 & \cellcolor{yellow!25}0.88 & \cellcolor{green!25}1.00 & \cellcolor{yellow!25}0.94 & \cellcolor{yellow!25}0.94 & \cellcolor{green!25}0.96 & \cellcolor{yellow!25}0.90 \\ \hline
3 & 3 & \cellcolor{yellow!25}0.94 & \cellcolor{green!25}0.98 & \cellcolor{green!25}0.98 & \cellcolor{green!25}0.96 & \cellcolor{yellow!25}0.94 & \cellcolor{green!25}0.98 & \cellcolor{yellow!25}0.94 & \cellcolor{green!25}0.98 & \cellcolor{green!25}0.98 & \cellcolor{yellow!25}0.90 & \cellcolor{yellow!25}0.94 & \cellcolor{green!25}1.00 & \cellcolor{green!25}0.98 & \cellcolor{yellow!25}0.94 & \cellcolor{yellow!25}0.92 \\ \hline
3 & 4 & \cellcolor{yellow!25}0.94 & \cellcolor{yellow!25}0.94 & \cellcolor{green!25}0.98 & \cellcolor{green!25}0.96 & \cellcolor{green!25}0.96 & \cellcolor{green!25}0.96 & \cellcolor{yellow!25}0.94 & \cellcolor{yellow!25}0.94 & \cellcolor{green!25}0.96 & \cellcolor{yellow!25}0.86 & \cellcolor{green!25}0.96 & \cellcolor{green!25}0.98 & \cellcolor{green!25}0.98 & \cellcolor{green!25}0.96 & \cellcolor{yellow!25}0.88 \\ \hline
4 & 2 & \cellcolor{yellow!25}0.82 & \cellcolor{yellow!25}0.84 & \cellcolor{green!25}0.98 & \cellcolor{yellow!25}0.90 & \cellcolor{green!25}0.96 & \cellcolor{green!25}0.98 & \cellcolor{green!25}0.98 & \cellcolor{green!25}0.96 & \cellcolor{green!25}0.96 & \cellcolor{yellow!25}0.86 & \cellcolor{green!25}1.00 & \cellcolor{green!25}1.00 & \cellcolor{yellow!25}0.94 & \cellcolor{green!25}0.96 & \cellcolor{yellow!25}0.92 \\ \hline
4 & 3 & \cellcolor{yellow!25}0.86 & \cellcolor{yellow!25}0.82 & \cellcolor{green!25}0.98 & \cellcolor{yellow!25}0.82 & \cellcolor{yellow!25}0.94 & \cellcolor{yellow!25}0.92 & \cellcolor{yellow!25}0.92 & \cellcolor{yellow!25}0.92 & \cellcolor{yellow!25}0.90 & \cellcolor{yellow!25}0.92 & \cellcolor{green!25}0.96 & \cellcolor{green!25}1.00 & \cellcolor{yellow!25}0.94 & \cellcolor{green!25}1.00 & \cellcolor{yellow!25}0.90 \\ \hline
4 & 4 & \cellcolor{yellow!25}0.90 & \cellcolor{yellow!25}0.90 & \cellcolor{green!25}0.96 & \cellcolor{yellow!25}0.92 & \cellcolor{green!25}0.96 & \cellcolor{yellow!25}0.90 & \cellcolor{yellow!25}0.94 & \cellcolor{yellow!25}0.94 & \cellcolor{yellow!25}0.88 & \cellcolor{yellow!25}0.92 & \cellcolor{green!25}1.00 & \cellcolor{green!25}0.98 & \cellcolor{green!25}0.98 & \cellcolor{yellow!25}0.94 & \cellcolor{yellow!25}0.92 \\ \hline
6 & 2 & \cellcolor{red!25}0.60 & \cellcolor{red!25}0.62 & \cellcolor{yellow!25}0.94 & \cellcolor{red!25}0.58 & \cellcolor{green!25}0.96 & \cellcolor{yellow!25}0.90 & \cellcolor{yellow!25}0.92 & \cellcolor{yellow!25}0.92 & \cellcolor{yellow!25}0.94 & \cellcolor{yellow!25}0.86 & \cellcolor{green!25}1.00 & \cellcolor{green!25}1.00 & \cellcolor{green!25}0.96 & \cellcolor{green!25}0.96 & \cellcolor{yellow!25}0.90 \\ \hline
6 & 3 & \cellcolor{red!25}0.72 & \cellcolor{red!25}0.74 & \cellcolor{green!25}0.96 & \cellcolor{red!25}0.68 & \cellcolor{green!25}0.96 & \cellcolor{yellow!25}0.80 & \cellcolor{yellow!25}0.92 & \cellcolor{yellow!25}0.82 & \cellcolor{red!25}0.78 & \cellcolor{yellow!25}0.90 & \cellcolor{green!25}0.96 & \cellcolor{green!25}1.00 & \cellcolor{green!25}0.96 & \cellcolor{green!25}1.00 & \cellcolor{yellow!25}0.90 \\ \hline
6 & 4 & \cellcolor{red!25}0.52 & \cellcolor{red!25}0.60 & \cellcolor{green!25}1.00 & \cellcolor{red!25}0.56 & \cellcolor{green!25}0.96 & \cellcolor{red!25}0.72 & \cellcolor{yellow!25}0.92 & \cellcolor{yellow!25}0.92 & \cellcolor{red!25}0.76 & \cellcolor{yellow!25}0.90 & \cellcolor{green!25}0.96 & \cellcolor{green!25}0.98 & \cellcolor{yellow!25}0.92 & \cellcolor{yellow!25}0.92 & \cellcolor{yellow!25}0.90 \\ \hline
8 & 2 & \cellcolor{red!25}0.62 & \cellcolor{green!25}0.98 & \cellcolor{green!25}0.98 & \cellcolor{red!25}0.60 & \cellcolor{yellow!25}0.92 & \cellcolor{red!25}0.76 & \cellcolor{green!25}0.98 & \cellcolor{red!25}0.78 & \cellcolor{red!25}0.76 & \cellcolor{yellow!25}0.88 & \cellcolor{green!25}0.96 & \cellcolor{green!25}0.98 & \cellcolor{green!25}0.98 & \cellcolor{yellow!25}0.94 & \cellcolor{yellow!25}0.94 \\ \hline
8 & 3 & \cellcolor{red!25}0.54 & \cellcolor{green!25}1.00 & \cellcolor{green!25}0.98 & \cellcolor{red!25}0.54 & \cellcolor{green!25}0.98 & \cellcolor{red!25}0.78 & \cellcolor{green!25}1.00 & \cellcolor{yellow!25}0.90 & \cellcolor{yellow!25}0.82 & \cellcolor{green!25}0.98 & \cellcolor{green!25}0.98 & \cellcolor{green!25}0.98 & \cellcolor{green!25}0.96 & \cellcolor{green!25}1.00 & \cellcolor{yellow!25}0.82 \\ \hline
8 & 4 & \cellcolor{red!25}0.54 & \cellcolor{green!25}1.00 & \cellcolor{green!25}0.96 & \cellcolor{red!25}0.54 & \cellcolor{yellow!25}0.94 & \cellcolor{red!25}0.68 & \cellcolor{green!25}1.00 & \cellcolor{green!25}1.00 & \cellcolor{red!25}0.74 & \cellcolor{green!25}0.96 & \cellcolor{yellow!25}0.92 & \cellcolor{green!25}1.00 & \cellcolor{yellow!25}0.92 & \cellcolor{yellow!25}0.94 & \cellcolor{yellow!25}0.94 \\ \hline
\end{tabular}%
}
\caption{Coverage of $95\ \%$ parameters credible intervals across all scenarios. Here, $s_f$ = spatial factor, $t_f$ = temporal factor, SE = standard error ($e$), $r_s$ = spatial range, $r_t$ = temporal range, Var = variance ($\sigma^2$), and Int. = intercept ($\beta_0$). Green denotes coverage $\geq 95\ \%$, yellow denotes $80\ \% \leq \text{coverage} < 95\ \%$, and red denotes $\text{coverage} < 80\ \%$.}
\label{tab:coverage-nonsep}
\end{table}

Spatio--temporal coverages for the non-separable model are reported in Table~\ref{tab:coverage2}. Values are empirical coverage probabilities (ECP) of nominal two-sided $95\%$ posterior credible intervals. Under weak autocorrelation, coverage tends to increase with $s_f$, peaking at $s_f=8$. With moderate and strong autocorrelation, coverage is more stable. To account for the effect of wider intervals, Table~\ref{tab:width2} reports mean widths.

\begin{table}[h!]
\small
\centering
\setlength{\tabcolsep}{4pt}
\begin{tabular}{l *{5}{c} *{5}{c} *{5}{c}}
\toprule
& \multicolumn{5}{c}{Weak} & \multicolumn{5}{c}{Moderate} & \multicolumn{5}{c}{Strong} \\
\cmidrule(lr){2-6}\cmidrule(lr){7-11}\cmidrule(lr){12-16}
$t_f \setminus s_f$ & 2 & 3 & 4 & 6 & 8 & 2 & 3 & 4 & 6 & 8 & 2 & 3 & 4 & 6 & 8 \\
\midrule
2 & 0.95 & 0.95 & 0.93 & 0.92 & 0.95 & 0.94 & 0.95 & 0.94 & 0.95 & 0.96 & 0.93 & 0.94 & 0.94 & 0.92 & 0.92 \\
3 & 0.94 & 0.94 & 0.92 & 0.95 & 0.96 & 0.94 & 0.94 & 0.95 & 0.96 & 0.96 & 0.93 & 0.93 & 0.93 & 0.92 & 0.94 \\
4 & 0.94 & 0.93 & 0.92 & 0.96 & 0.98 & 0.94 & 0.95 & 0.95 & 0.93 & 0.96 & 0.94 & 0.93 & 0.92 & 0.91 & 0.94 \\
\bottomrule
\end{tabular}
\caption{Empirical coverage probability (ECP) of nominal two-sided $95\%$ posterior credible intervals. Rows index the temporal factor $t_f$; columns the spatial factor $s_f$ for the non--separable model.}
\label{tab:coverage2}
\end{table}

\begin{table}[h!]
\small
\centering
\setlength{\tabcolsep}{4pt}
\begin{tabular}{l *{5}{c} *{5}{c} *{5}{c}}
\toprule
& \multicolumn{5}{c}{Weak} & \multicolumn{5}{c}{Moderate} & \multicolumn{5}{c}{Strong} \\
\cmidrule(lr){2-6}\cmidrule(lr){7-11}\cmidrule(lr){12-16}
$t_f \setminus s_f$ & 2 & 3 & 4 & 6 & 8 & 2 & 3 & 4 & 6 & 8 & 2 & 3 & 4 & 6 & 8 \\
\midrule
2 & 0.59 & 0.66 & 0.75 & 1.00 & 1.10 & 0.59 & 0.69 & 0.78 & 1.04 & 1.23 & 0.52 & 0.63 & 0.71 & 0.84 & 0.93 \\
3 & 0.65 & 0.71 & 0.87 & 1.07 & 1.22 & 0.66 & 0.76 & 0.94 & 1.27 & 1.27 & 0.55 & 0.65 & 0.75 & 0.89 & 1.01 \\
4 & 0.69 & 0.75 & 0.81 & 1.29 & 1.30 & 0.71 & 0.84 & 1.09 & 1.23 & 1.31 & 0.60 & 0.69 & 0.76 & 0.96 & 1.14 \\
\bottomrule
\end{tabular}
\caption{Mean width of the two-sided $95\%$ posterior credible intervals, averaged over space and time. Rows index $t_f$, columns index $s_f$ within each temporal-autocorrelation configuration for the non--separable model.}
\label{tab:width2}
\end{table}

Together, the results show that higher coverage at large $s_f$ under weak autocorrelation comes at the cost of wider intervals. In contrast, moderate and strong autocorrelation yields narrower intervals for comparable $(s_f,t_f)$, indicating more efficient borrowing of information.

\section{Spatio--temporal disaggregation of Aerosol Optical Depth at 550 nm in India}
\label{sec:app}

We utilize our spatio-temporal modeling approach to disaggregate satellite derived aerosol optical depth (AOD) at 550 $\mathrm{nm}$, a common indicator of air pollution \cite{donkelaaretal2010}.
We downscale AOD in India, chosen for its elevated levels in early 2024 and its varied topography, including both high elevations and coastal regions.

Our objective is to generate high-resolution spatio–-temporal estimates of AOD, which are critical for exposure assessment and policy evaluation. To achieve this, we model AOD within a flexible spatio–-temporal framework, incorporating altitude as a key covariate to capture topographic relationships with aerosol distribution. This approach yields more detailed AOD maps that better capture elevation-related pollution patterns, improving their utility for informed decision-making.

We consider AOD at 550 $\mathrm{nm}$ from the Atmospheric Composition Reanalysis $4$ (EAC4) of the European Centre for Medium‐Range Weather Forecasts (ECMWF). EAC4 combines model data with global observations into a consistent dataset based on the laws of physics and chemistry \cite{inness2019cams}. The dataset is available from 2003 onward at 3-hourly resolution on a global $0.75^\circ \times 0.75^\circ$ grid. For elevation, we use the SRTM30$+$ Version $11$ land‐surface digital elevation model (DEM), produced by the Scripps Institution of Oceanography (SIO), which provides global land elevation at approximately $ 1\mathrm{km}$ resolution on a regular latitude–-longitude grid \cite{SRTM30plus_v11_land_PacIOOS_2014}.

The analysis uses the first 5 days of January $2024$ data sampled every $3$ hours on a $0.75^{\circ}\times 0.75^{\circ}$ grid comprising $34$ longitudes by $37$ latitudes ($34\times 37 = 1{,}258$ locations). The spatio–-temporal disaggregation increases the temporal resolution from $3\mathrm{h}$ to $1\mathrm{h}$ and the spatial resolution from $0.75^{\circ}$ to $0.25^{\circ}$, expanding the spatial locations from $1{,}258$ to  $1{,}258 \times 9 = 11{,}322$ and the temporal points from $40$ to $40\times 3 = 120$. The model used here follows a similar framework than the simulation study (see Equations~\eqref{equ:cont-eq} and \eqref{equ:agg-1}), with elevation added as an explicit covariate:

\begin{equation}
    AOD (\textbf{s},t) = \beta_0 + \beta_1 \cdot \text{elevation}(\textbf{s}) + z(\textbf{s},t) + e(\textbf{s},t).
\end{equation}

\begin{figure}[!h]
\centering
\setlength{\tabcolsep}{2pt}
\setlength{\arrayrulewidth}{0.6pt}
\renewcommand{\arraystretch}{0}
\begin{tabular}{@{}ccc|c@{}}
\includegraphics[width=.24\textwidth]{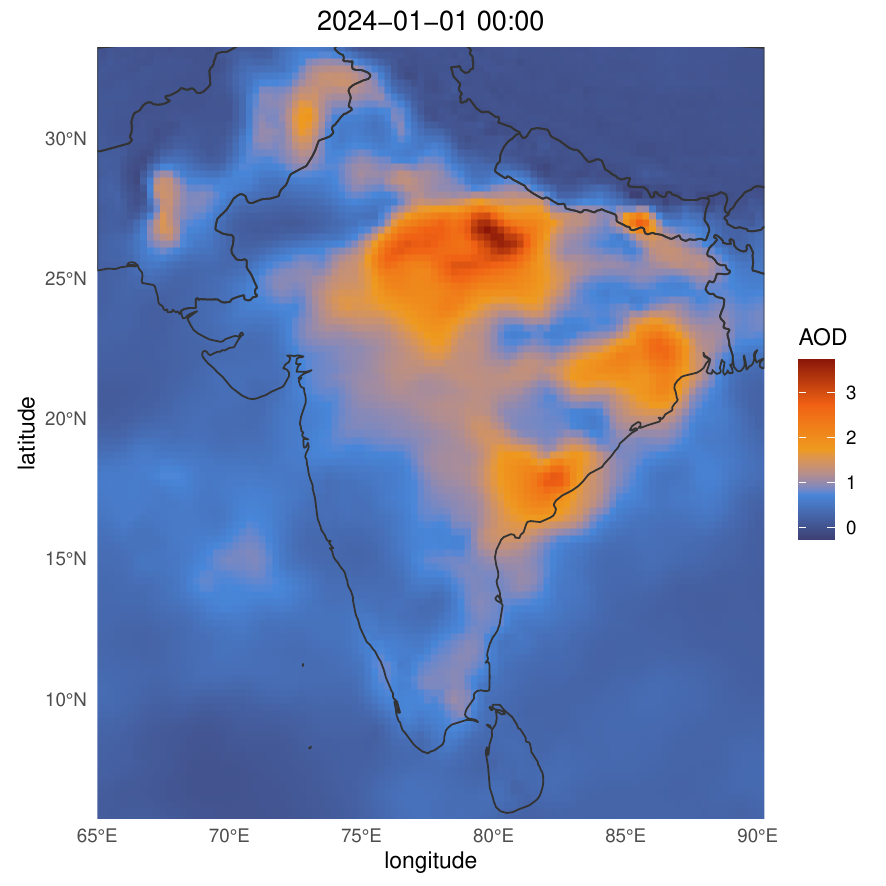} &
\includegraphics[width=.24\textwidth]{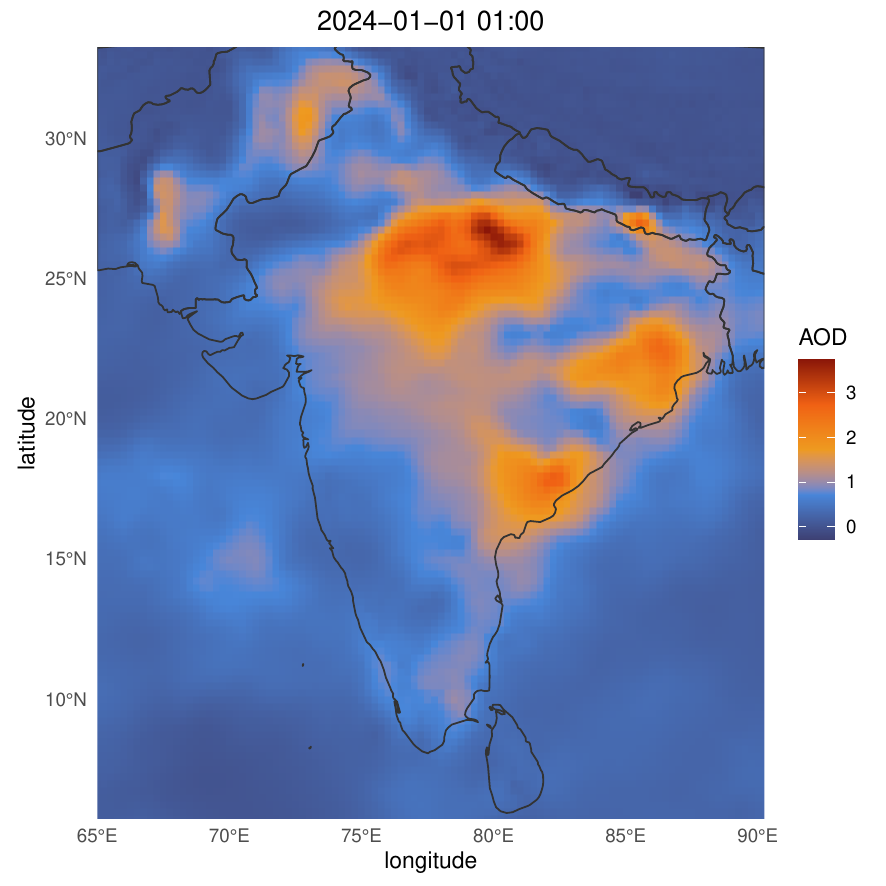} &
\includegraphics[width=.24\textwidth]{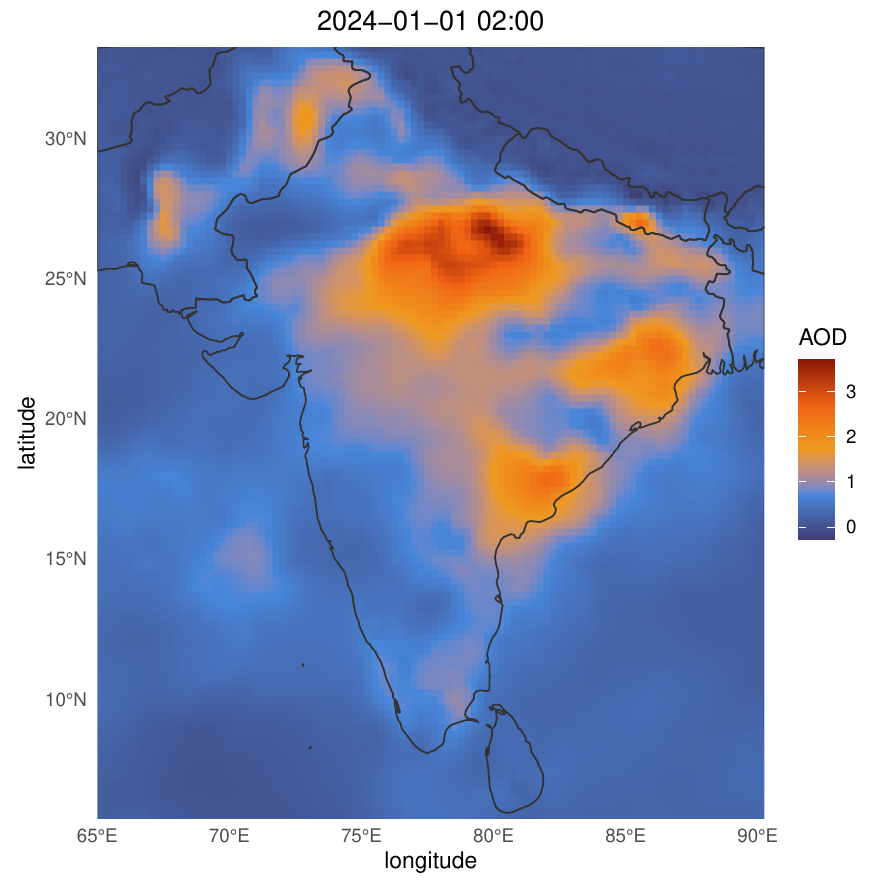} &
\includegraphics[width=.24\textwidth]{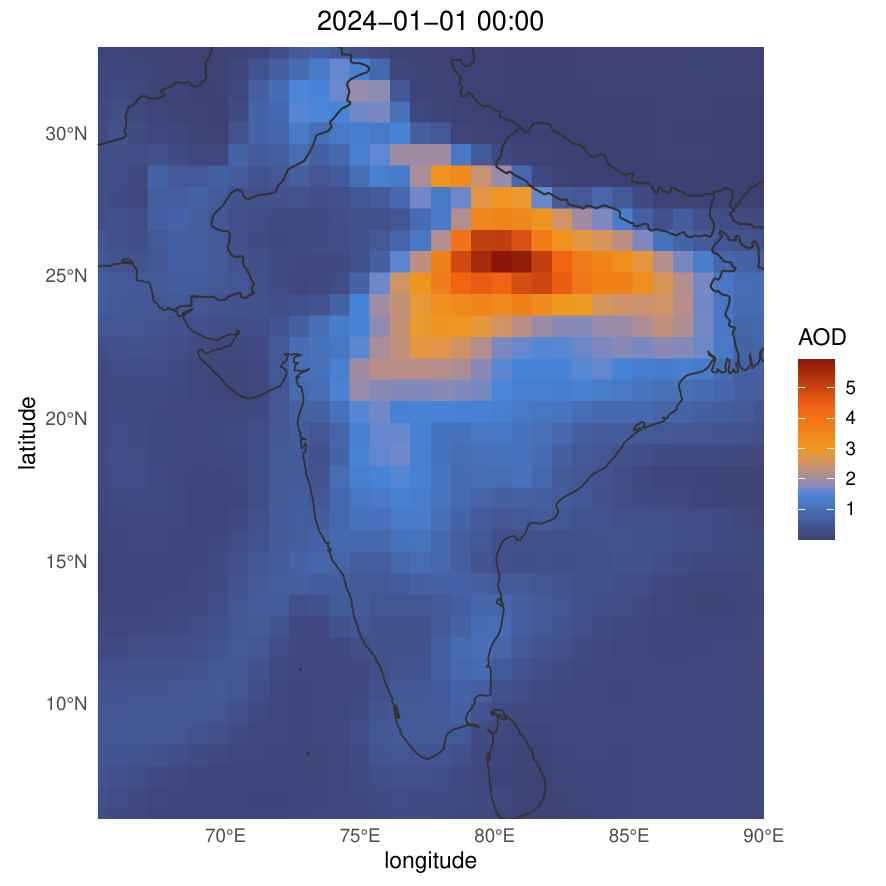} \\
\vspace{0.3cm} \\
\includegraphics[width=.24\textwidth]{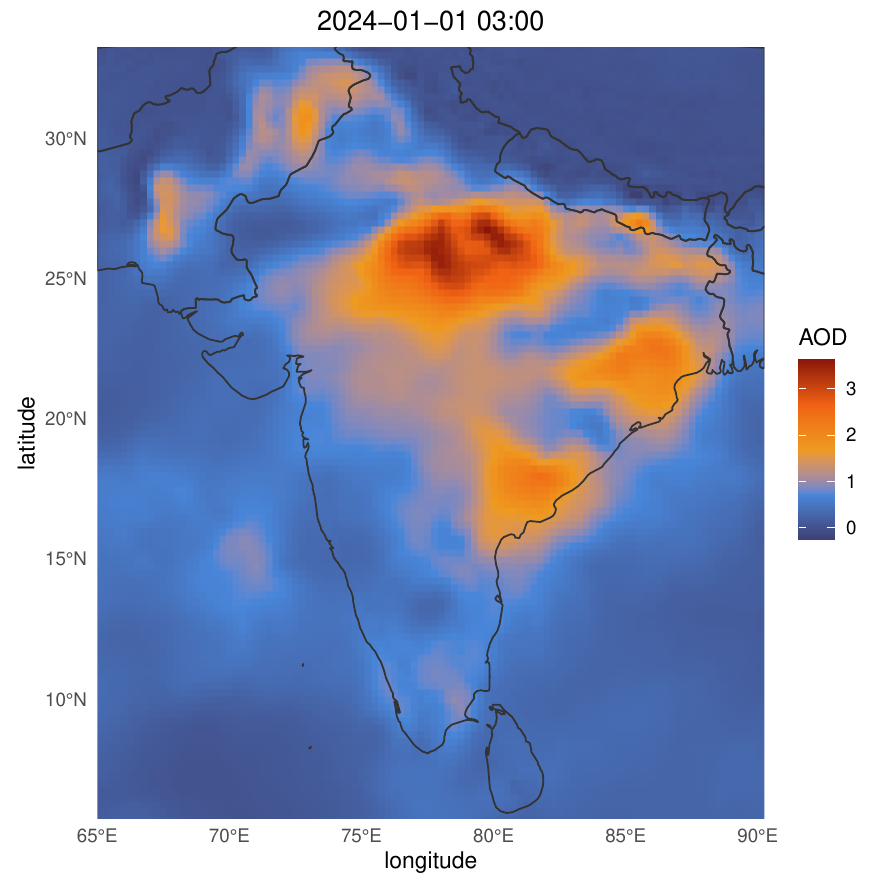} &
\includegraphics[width=.24\textwidth]{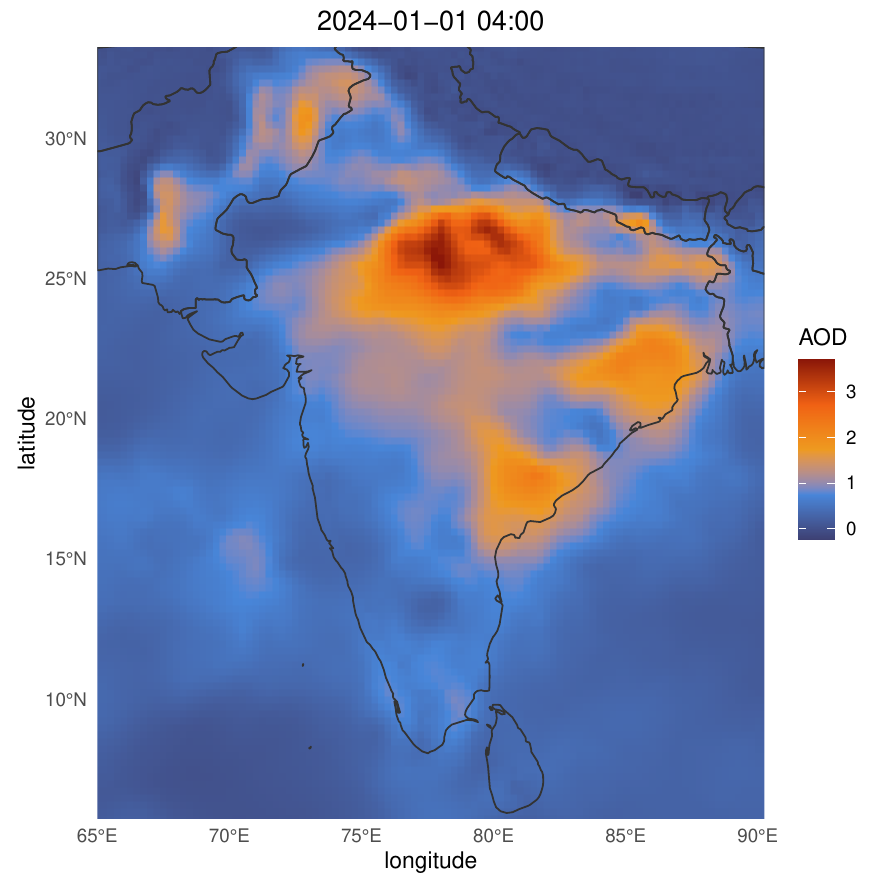} &
\includegraphics[width=.24\textwidth]{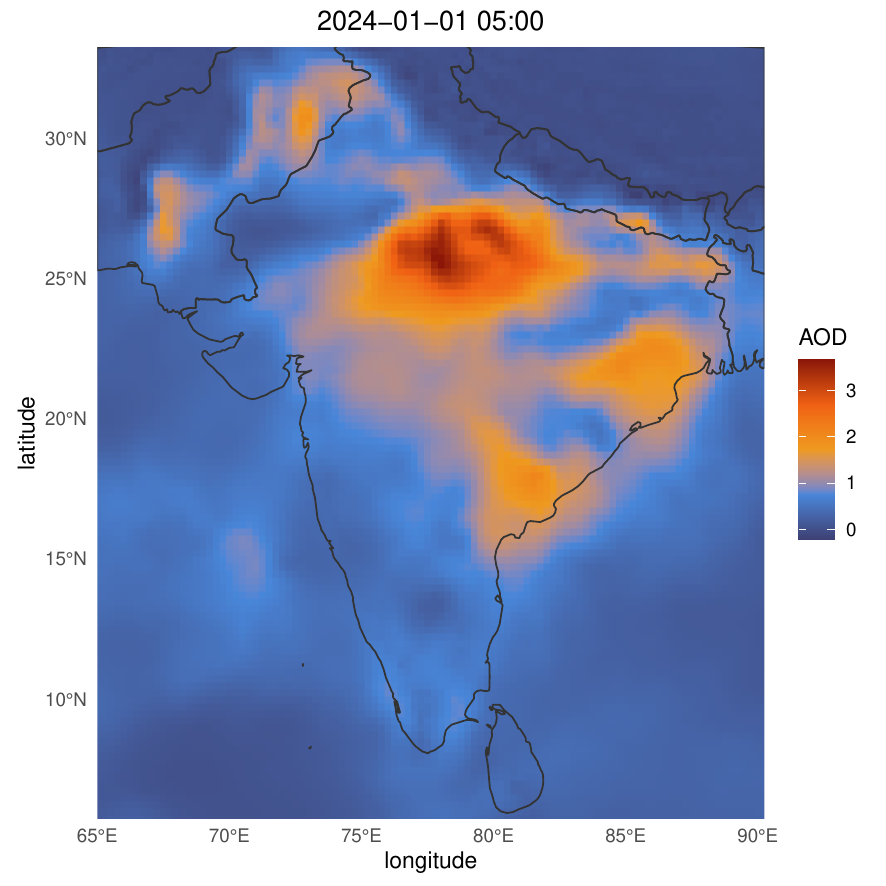} &
\includegraphics[width=.24\textwidth]{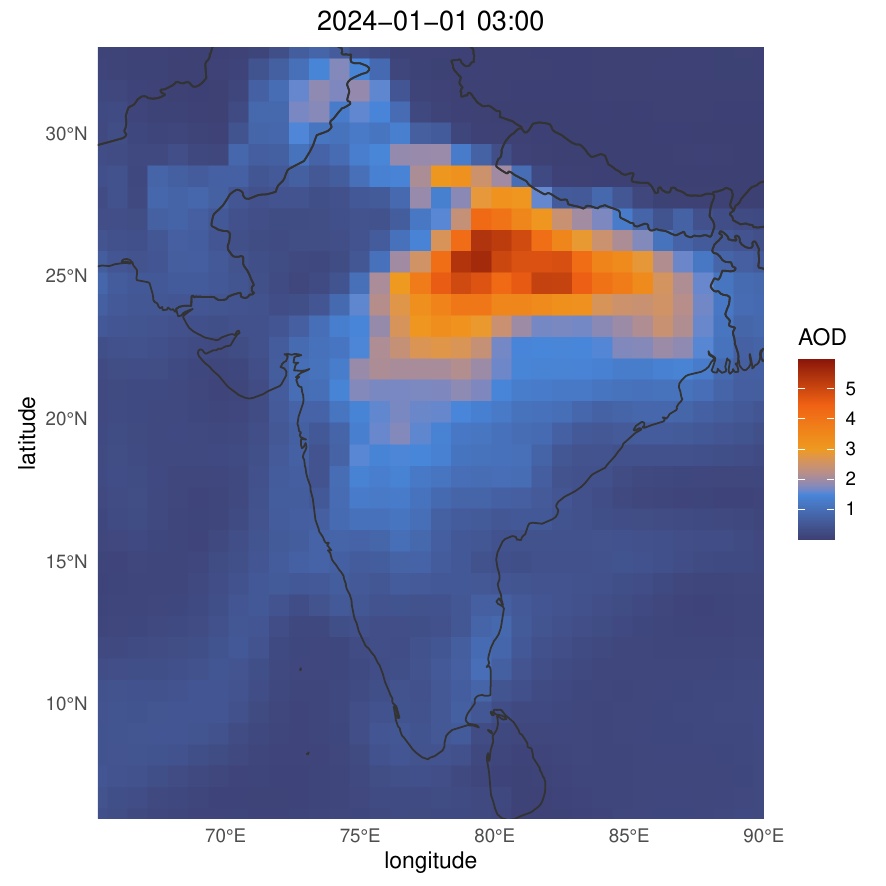} 
\end{tabular}
\caption{Posterior–mean aerosol optical depth (AOD) over India on 1~January~2024 for the first six hours (left three columns), alongside the original data at the aggregated resolution for the same day at 00{:}00 and 03{:}00 (right column).}
\label{fig:app-res}
\end{figure}

In Figure~\ref{fig:app-res}, we show posterior means from our spatio–-temporal disaggregation applied to AOD over India on 1~January~2024. The left three columns display the first six hours of the day at hourly resolution, while the right column shows the original data at $00{:}00$ and $03{:}00$. The figure illustrates that the spatio--temporal disaggregation preserves the scale and variability of the source data while providing a spatially smoother representation that captures the spatio--temporal trend that smoothly changes according to the distribution of this pollutant. Our model also offers hourly estimations, providing a smoother representation over time as observations transition from being sampled 3 hours apart to being taken every hour. An animated plot showing the variation of AOD in space and time can be seen in the GitHub repository of this paper \url{https://github.com/fravellaneda/spatio-temporal-disaggregation/aod-distribution-junary-1-5-2024.gif}.

\begin{figure}
\centering
\includegraphics[width=1\linewidth]{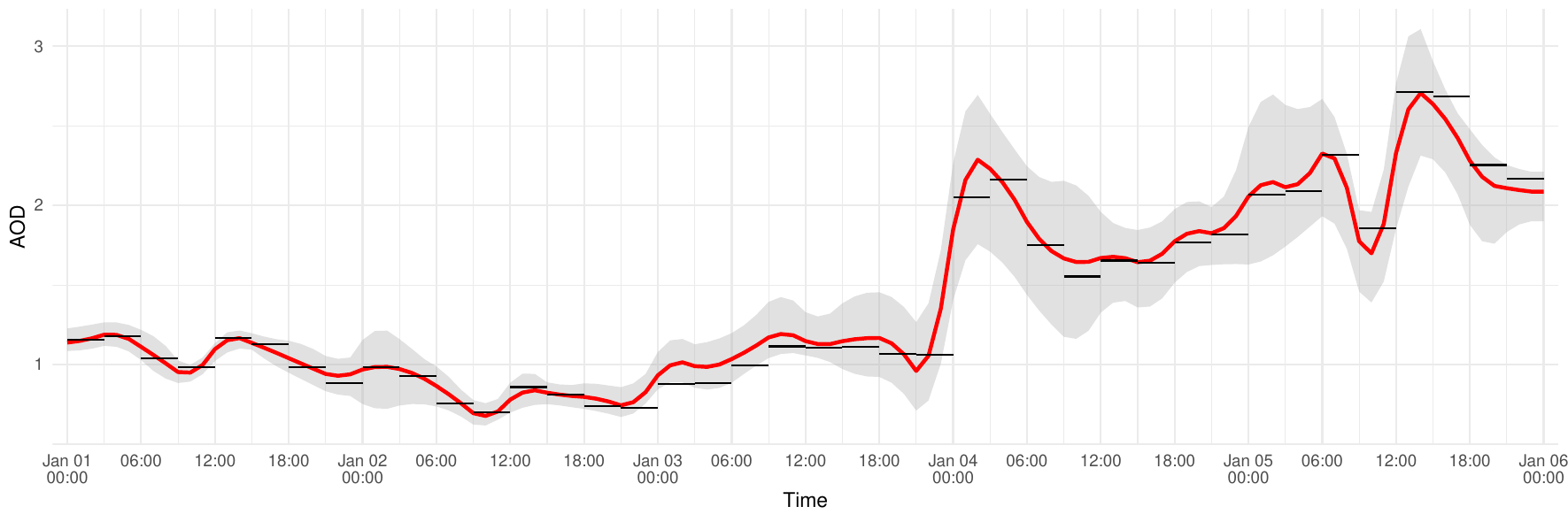}
\caption{Observed and predicted AOD concentrations for the grid cell containing New Delhi. Black segments indicate the coarse-resolution observations ($0.75^\circ$, 3 hours). The red line shows the average prediction from the nine finer-resolution cells, and the shaded area marks the minimum and maximum predicted values across those cells.}
\label{fig:delhi}
\end{figure}

In Figure~\ref{fig:delhi}, we show the observed and predicted AOD values for the grid cell containing New Delhi, the capital of India. The black segments represent the original coarse-resolution observations ($0.75^\circ$ in space and 3 hours in time). The red line corresponds to the average AOD concentration estimated by our model across the nine finer-resolution cells (a $3 \times 3$ grid) within the original coarse cell. The shaded area denotes the minimum and maximum values among these finer cells. This figure demonstrates that the model not only captures the overall temporal trend but also provides a smoother and more continuous representation of the variability, here resolved at an hourly scale, compared to the coarse observations.

In Table~\ref{tab:app-params}, we present the summary of the fixed parameters and hyperparameters used in the model. For the fixed parameters, the intercept gives us the measure of the mean AOD level in India which is estimated as $0.62$. The other fixed effect, the elevation, is captured by the coefficient $\beta_1$ that quantifies the linear relationship between elevation and AOD, in this case, a elevation of $1\mathrm{km}$, reduces AOD by about $0.08$, with a $95\%$ credible interval entirely below zero which confirm us that elevation is a good predictor of the AOD levels.

\begin{table}[h!]
\centering
\begin{tabular}{|c|r|r|r|r|}
\hline
\textbf{Parameter} & \textbf{Mean} & \textbf{0.025 Quantile} & \textbf{0.5 Quantile} & \textbf{0.975 Quantile} \\ \hline
$\beta_0$ & 0.62 & -1.65 & 0.62 & 2.89 \\ \hline
$\beta_1$ & -0.08 & -0.10 & -0.08 & -0.06 \\ \hline
$\tau$ & 130448.32 & 91731.39 & 128488.58 & 180085.62 \\ \hline
$r_s$ & 15.17 & 13.28 & 15.07 & 17.51 \\ \hline
$r_t$ & 6108.16 & 4640.17 & 6001.74 & 8132.78 \\ \hline
$\sigma$ & 1.29 & 1.13 & 1.28 & 1.49 \\ \hline
\end{tabular}
\vspace{0.2cm}
\caption{Summary of fixed effects intercept $\beta_0$, elevation $\beta_1$, and hyperparameters.
$\tau$ is the precision of the spatio--temporal Gaussian noise; $r_s$ and $r_t$ are the spatial and temporal range, and $\sigma$ is the variance of the spatio--temporal non--separable Gaussian process.}
\label{tab:app-params}
\end{table}

For the random parameters of the Gaussian process in Table~\ref{tab:app-params}, we note that the large mean temporal range, quantified by $r_t \approx 6108.16$, indicates strong temporal autocorrelation. This allows the model to borrow information across nearby time points, which enhances the quality of the spatio–-temporal disaggregation. Additionally, the high mean value of the precision for the spatio–-temporal Gaussian noise ($\tau = 130448.32$) and the moderate mean value of the variance for the non--separable Gaussian process ($\sigma = 1.29$) indicate that most of the variability is attributed to the latent spatio–-temporal process $z(\mathbf{s},t)$ rather than the unstructured random error. 

\begin{figure}[!h]
\centering
\setlength{\tabcolsep}{2pt}
\setlength{\arrayrulewidth}{0.6pt}
\renewcommand{\arraystretch}{0}
\begin{tabular}{@{}ccc@{}}
\includegraphics[width=.32\textwidth]{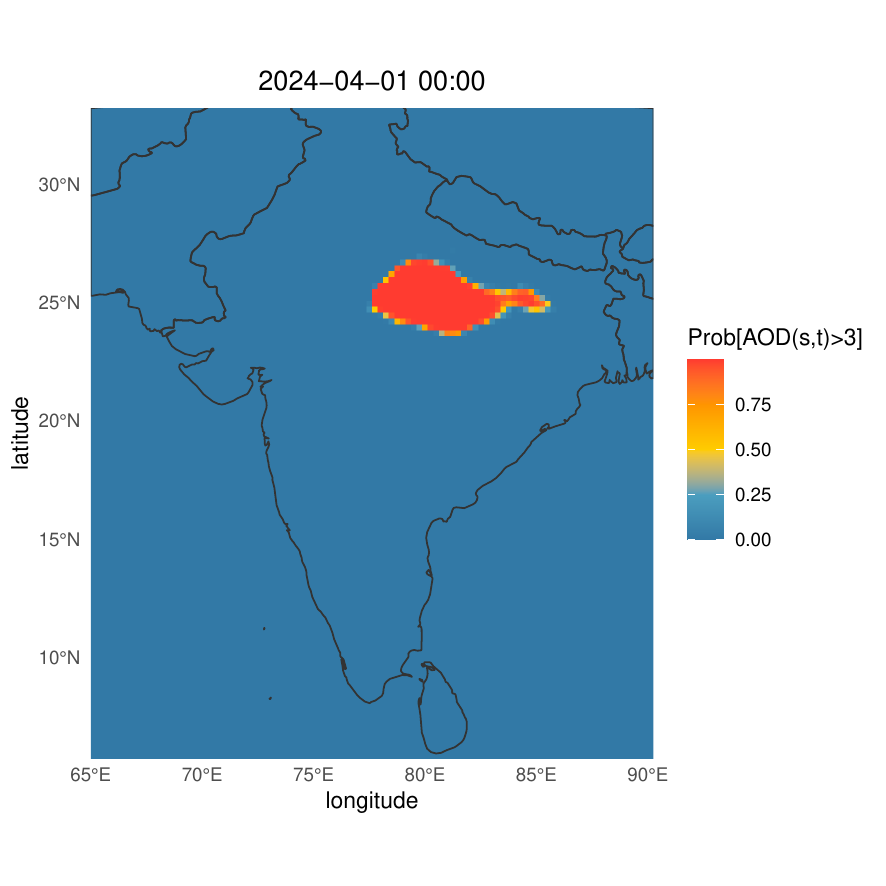} &
\includegraphics[width=.32\textwidth]{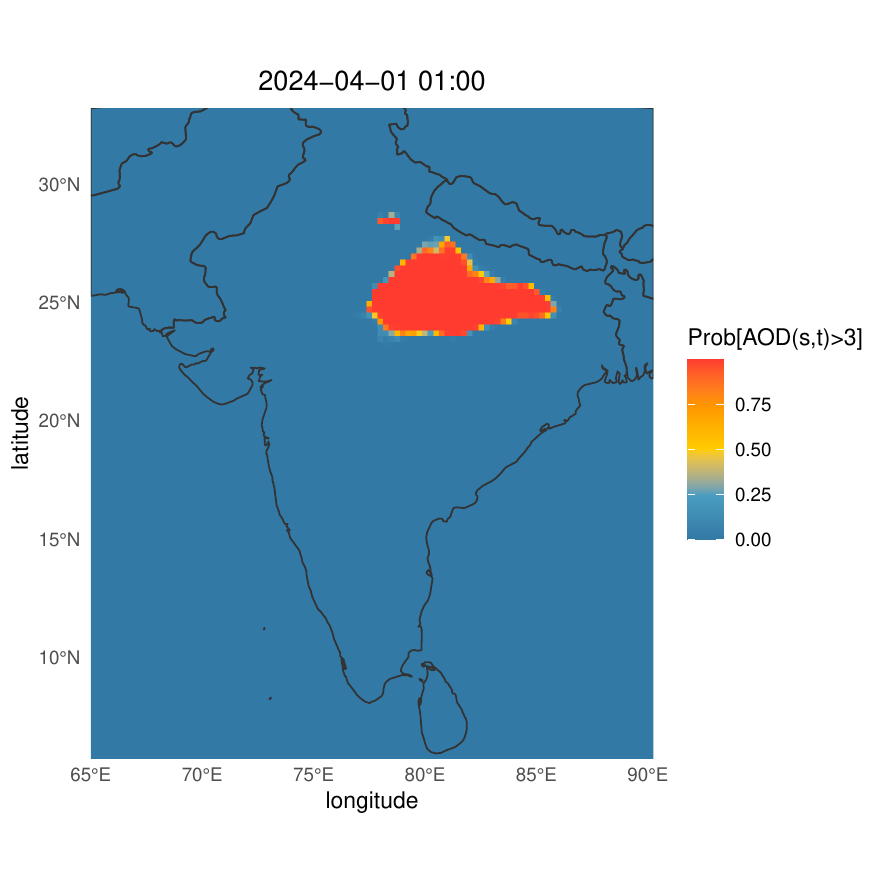} &
\includegraphics[width=.32\textwidth]{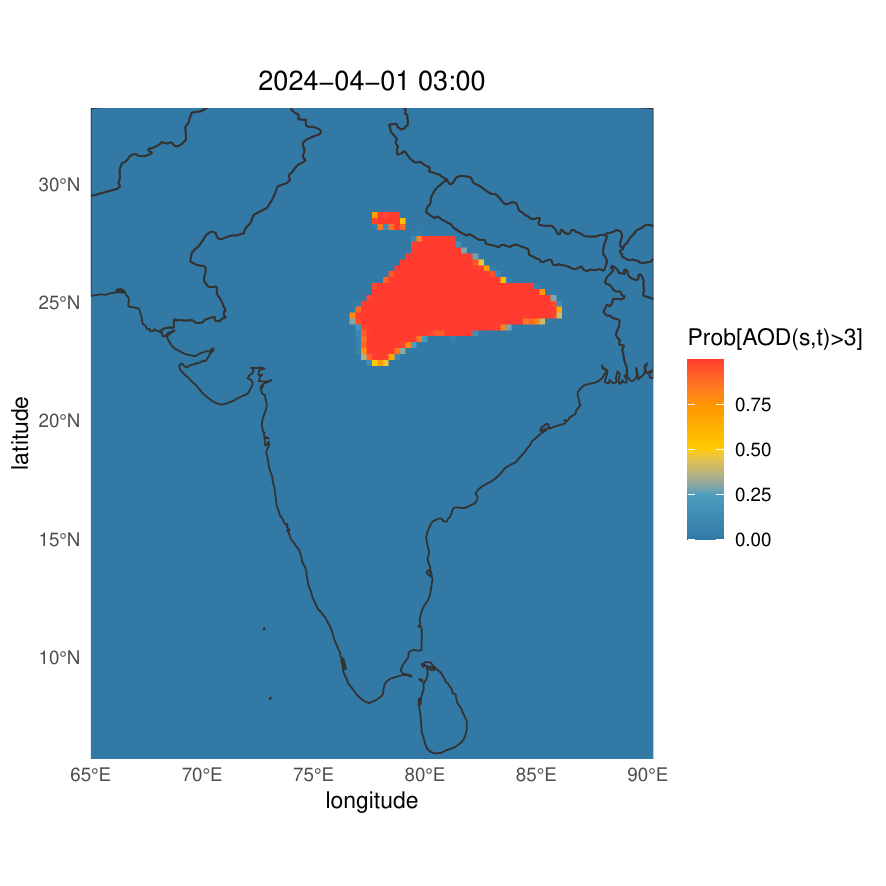}  \\
\includegraphics[width=.32\textwidth]{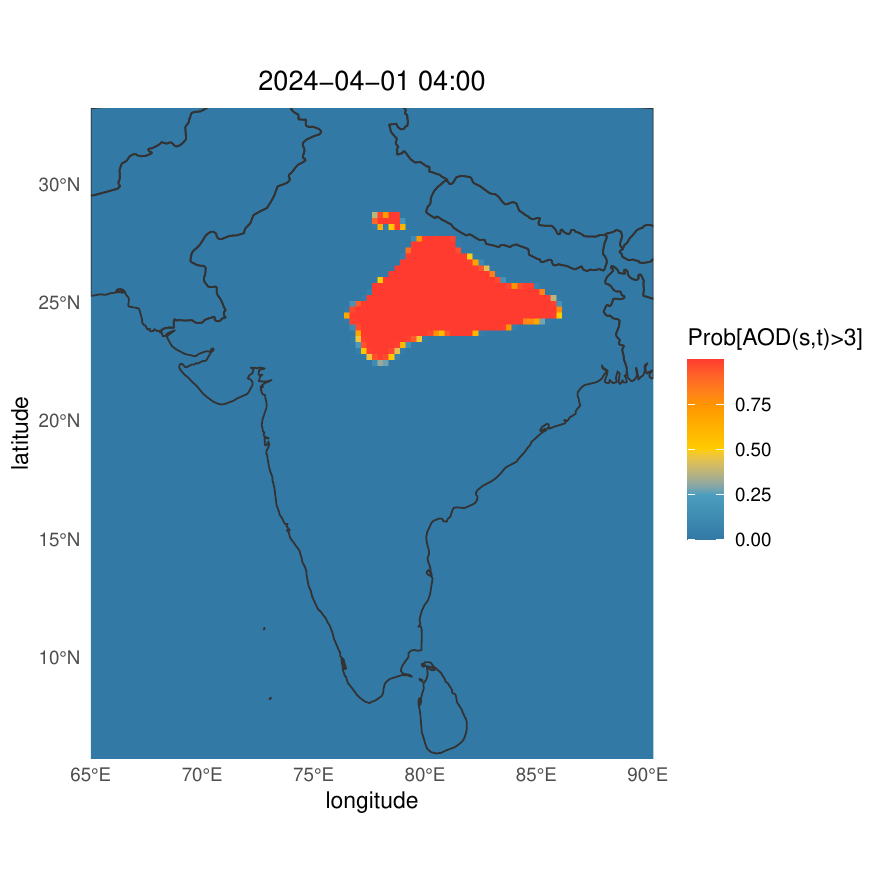} &
\includegraphics[width=.32\textwidth]{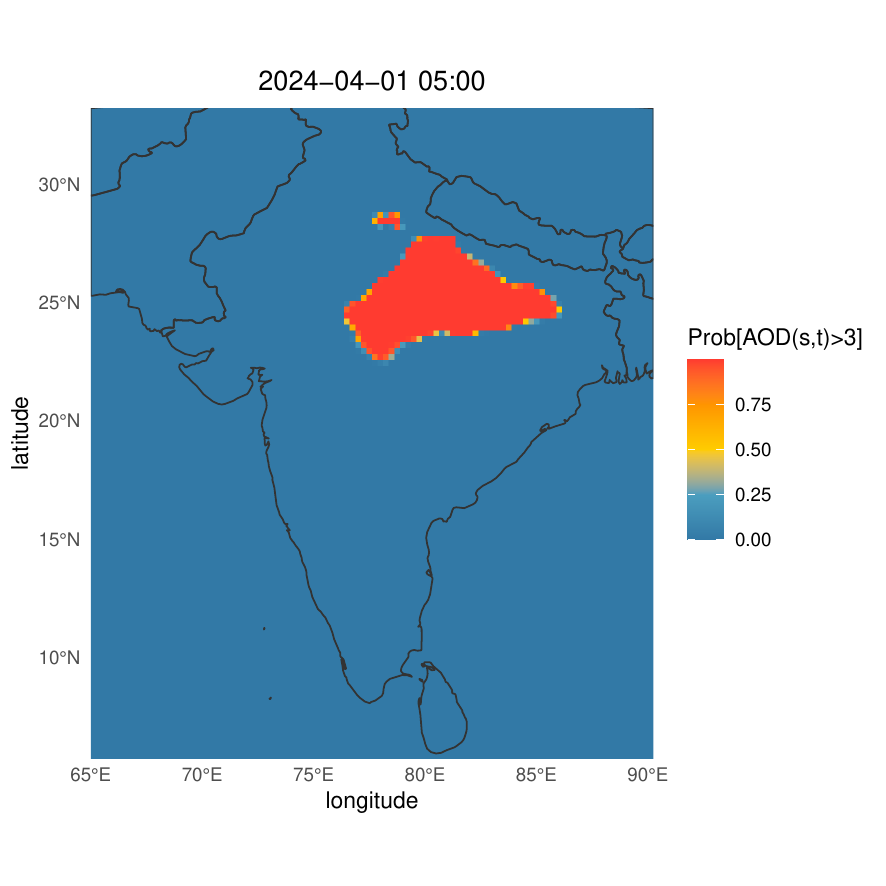} &
\includegraphics[width=.32\textwidth]{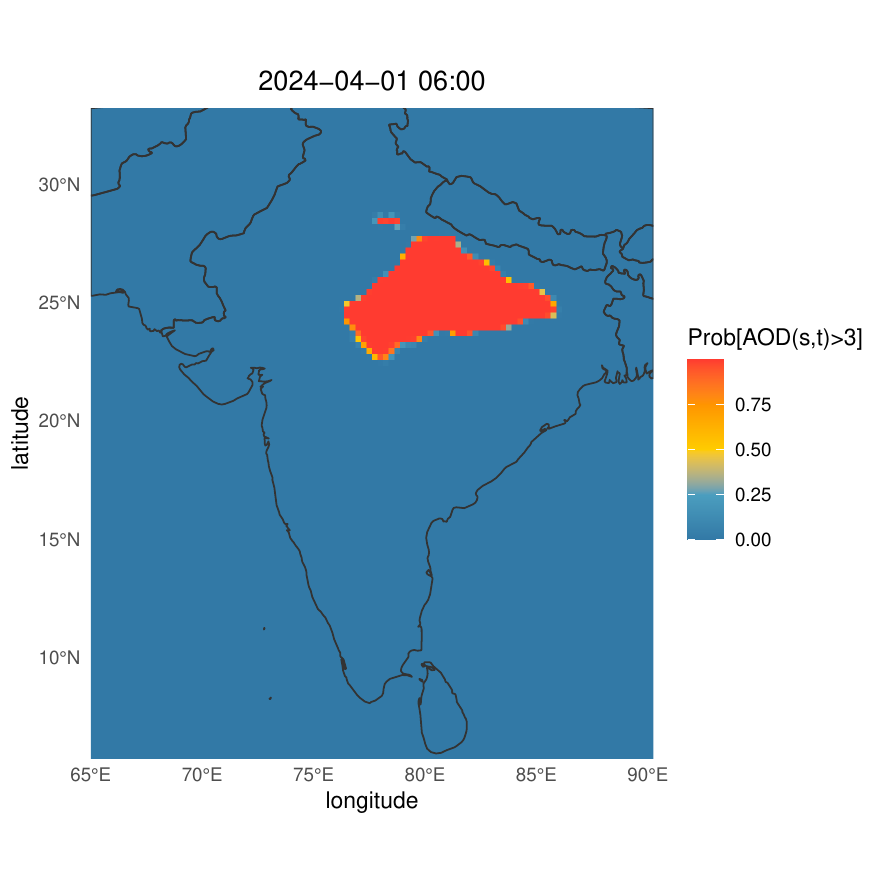} 
\end{tabular}
\caption{Posterior exceedance probability $\Pr\{\widehat{\mathrm{AOD}_{550}}>3\}$ over India for six consecutive hours on 4~January~2024.}
\label{fig:app-exc}
\end{figure}

Finally, we report exceedance probabilities for AOD at $550\,\mathrm{nm}$. Because AOD is dimensionless and not governed by health-based standards, the choice of threshold depends on the country and the location. Following NASA EOSDIS Earthdata’s interpretive guidance, where values greater than 3 indicate aerosol loads dense enough to obscure the Sun, we set the exceedance level to $3.0$ \cite{NASA_Earthdata_AOD_Thickness_2025}. Figure~\ref{fig:app-exc} maps the posterior exceedance probability $\Pr\{\widehat{\mathrm{AOD}_{550}}>3\}$ for six consecutive hourly windows on 4 January 2024 over India.

\section{Conclusion and discussion}
\label{sec:con}

In this work, we introduce a novel spatio--temporal disaggregation model that leverages a non--separable Gaussian process to capture complex dependencies across space and time. In this approach, we assume that there exists a spatio--temporal latent field, in which the observed data are the average observation in space and time. This formulation enables more flexible modeling of real-world phenomena, particularly in settings where separable structures may be overly restrictive. To ensure computational feasibility, especially when dealing with a high amount of data, we implement the model using the \texttt{R-INLA} and \texttt{INLAspacetime} framework, enabling fast and scalable Bayesian inference.

We conducted a simulation study to evaluate the performance of our spatio–-temporal disaggregation model in recovering the latent field from aggregated observations. The study considers two cases: one with a separable covariance structure and another with a non--separable structure. In the separable case, we compare our model with a traditional areal approach. Results show that our disaggregation model outperforms the areal separable alternative in most scenarios, yielding more accurate approximations of the underlying latent process. Across both frameworks, we observe that the model’s ability to recover the latent field improves as temporal autocorrelation increases, that is, when the underlying process exhibits strong temporal dependence, the disaggregation performs more effectively.

To demonstrate the efficiency of our methodology, we apply it to model Aerosol Optical Depth (AOD) at $550\ \mathrm{nm}$ over India. The original dataset, obtained from the Atmospheric Composition Reanalysis~4 (EAC4) of the European Centre for Medium-Range Weather Forecasts (ECMWF), provides data at a spatial resolution of $0.75^{\circ}$ in both longitude and latitude, and a temporal resolution of $3$ hours. Our model achieves a spatio–-temporal disaggregation to a finer resolution of $0.25^{\circ}$ spatially and $1$ hour temporally, representing a $27$-fold increase in resolution. In addition, the model allows for the inclusion of covariates such as elevation, which can enhance predictive performance and improve interpretability by accounting for known environmental influences on AOD levels.

Previous studies have shown that the interdependence among pollutants such as PM2.5, PM10, and ozone can increase the explanatory power of disaggregation models \cite{avellaneda2025multivariatedisaggregationmodelingair}. Building on this idea, a natural extension of our work would be to develop a multivariate spatio--temporal disaggregation model. Some initial models for spatial disaggregation with spatio--temporal data have been proposed \cite{gaedkemerzhäuser2025acceleratedspatiotemporalbayesianmodeling}, and these ideas can be extended to achieve spatio--temporal disaggregation with more than one variable. In this multivariate setting, another promising direction is to handle variables measured at different spatial or temporal resolutions. Instead of treating each variable separately, they can be modeled jointly to achieve a consistent and coherent disaggregation across both space and time, even when the data sources differ in resolution. A further practical step is to combine heterogeneous support variables measured at different spatial (point vs. area) or temporal (hourly vs. daily) resolutions \cite{zhongetal25}.

In addition, we note that for optimal performance, this method should be applied to a dataset with strong temporal autocorrelation, which enables the effective borrowing of information across time and improves interpolation between observations. When temporal dependence is weak, temporal borrowing diminishes, leading to larger posterior uncertainty and limited gains from temporal refinement.

In summary, this paper introduces a robust and novel spatio--temporal disaggregation model based on Gaussian processes with a non--separable covariance structure. This approach allows us to capture complex spatio–-temporal patterns that separable models cannot.
We applied the model to air pollution data, specifically AOD over India, and showed that it effectively enhances temporal and spatial resolution. Importantly, the proposed framework is versatile and can be applied to a wide range of spatio–-temporal variables \cite{moragaandbaker22}. For example, the proposed model can be applied to obtain high-resolution estimates of land-surface temperature, precipitation, soil moisture (SMAP), sea-surface temperature, and vegetation indices (NDVI/EVI).
This is especially valuable in situations where observations are only available in aggregated form across space and time, but the aim is to recover the underlying process at a finer resolution to better support decision-making.

\clearpage

\bibliographystyle{apalike}
\bibliography{biblio.bib}       

\clearpage

\begin{appendices}

\section{Simulation Results}
\label{sec:appendix-sim-res}

\begin{figure}[h!]
  \centering
  \begin{subfigure}[c]{0.31\textwidth}
    \includegraphics[width=0.9\linewidth]{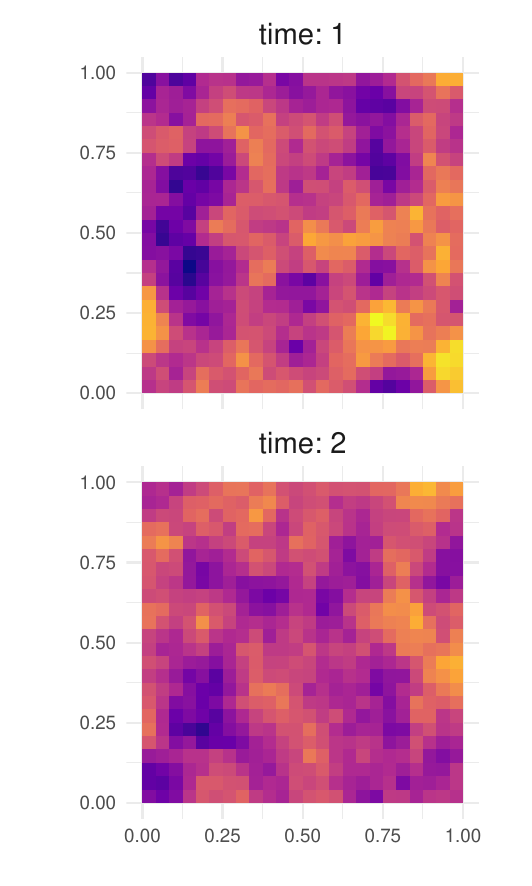}
    \caption{$W(\mathbf{s},t)$}
    \label{fig:app-sep-w-1}
  \end{subfigure}
  \hfill
  \begin{subfigure}[c]{0.31\textwidth}
    \includegraphics[width=0.9\linewidth]{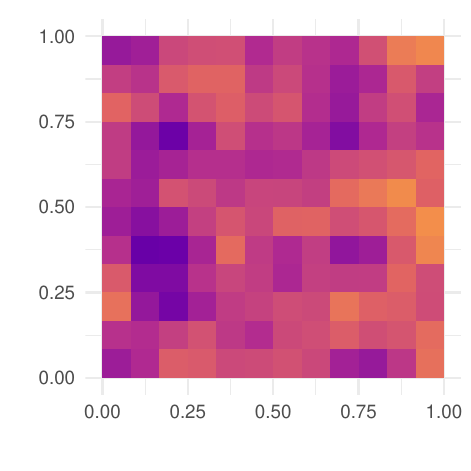}
    \caption{$W(R_{ij})$}
    \label{fig:app-sep-w-2}
  \end{subfigure}
  \hfill
  \begin{subfigure}[c]{0.31\textwidth}
    \includegraphics[width=0.9\linewidth]{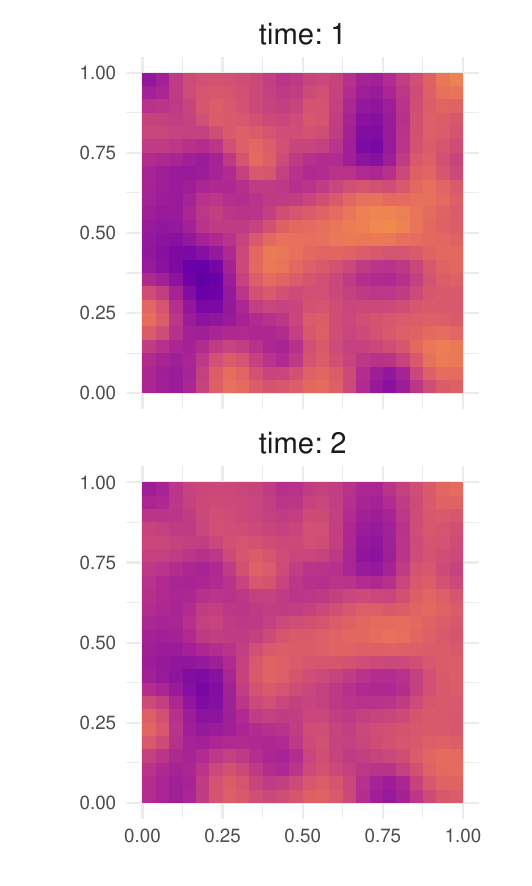}
    \caption{$\widehat{W}(\mathbf{s},t)$}
    \label{fig:app-sep-w-3}
  \end{subfigure}
  \caption{(a) Realization of the continuous process $W(\mathbf{s},t)$ in the \textit{separable} setting with \textbf{weak temporal autocorrelation}, (b) corresponding aggregated observations $W(R_{ij})$ for spatial factor $s_f =2$ and temporal factor $t_f=2$, and (c) the reconstructed process from the disaggregation model $\widehat{W}(\mathbf{s},t)$.}
  \label{fig:appendix-sep-1}
\end{figure}

\begin{figure}[h!]
  \centering
  \begin{subfigure}[c]{0.31\textwidth}
    \includegraphics[width=0.9\linewidth]{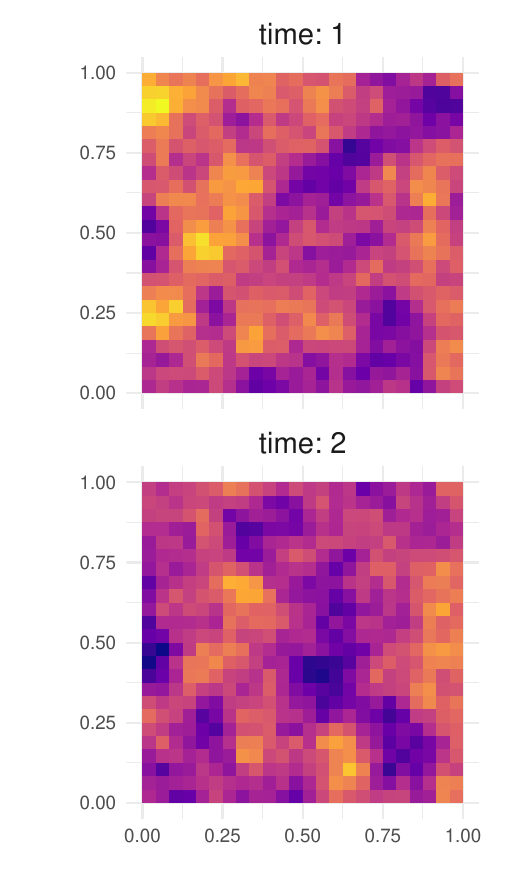}
    \caption{$W(\mathbf{s},t)$}
    \label{fig:app-sep-m-1}
  \end{subfigure}
  \hfill
  \begin{subfigure}[c]{0.31\textwidth}
    \includegraphics[width=0.9\linewidth]{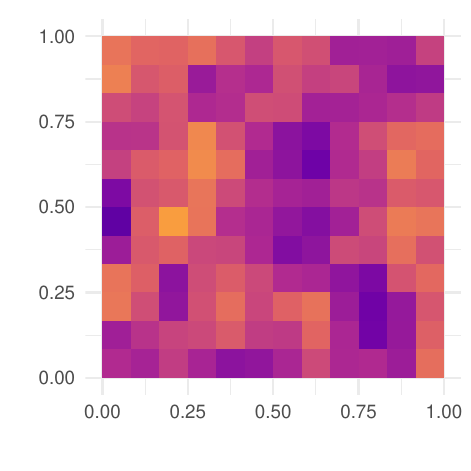}
    \caption{$W(R_{ij})$}
    \label{fig:app-sep-m-2}
  \end{subfigure}
  \hfill
  \begin{subfigure}[c]{0.31\textwidth}
    \includegraphics[width=0.9\linewidth]{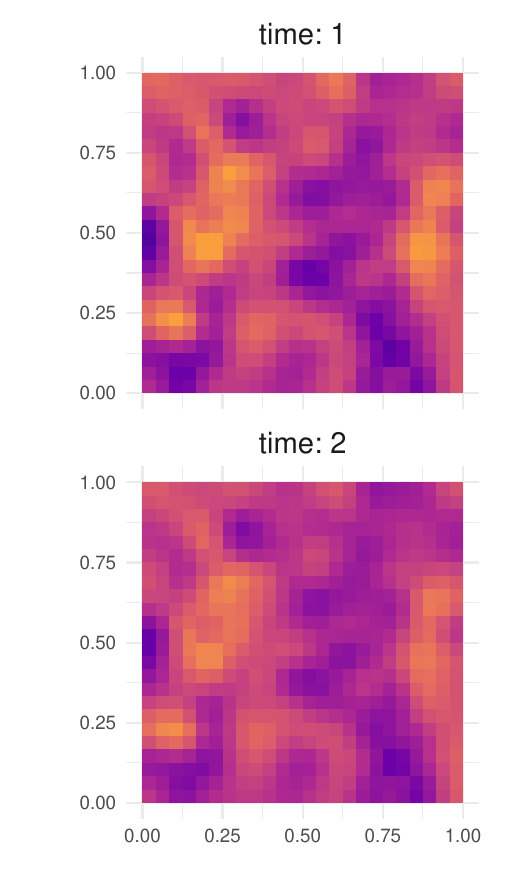}
    \caption{$\widehat{W}(\mathbf{s},t)$}
    \label{fig:app-sep-m-3}
  \end{subfigure}
  \caption{(a) Realization of the continuous process $W(\mathbf{s},t)$ in the \textit{separable} setting with \textbf{moderate temporal autocorrelation}, (b) corresponding aggregated observations $W(R_{ij})$ for spatial factor $s_f =2$ and temporal factor $t_f=2$, and (c) the reconstructed process from the disaggregation model $\widehat{W}(\mathbf{s},t)$.}
  \label{fig:appendix-sep-2}
\end{figure}

\begin{figure}[h!]
  \centering
  \begin{subfigure}[c]{0.31\textwidth}
    \includegraphics[width=0.9\linewidth]{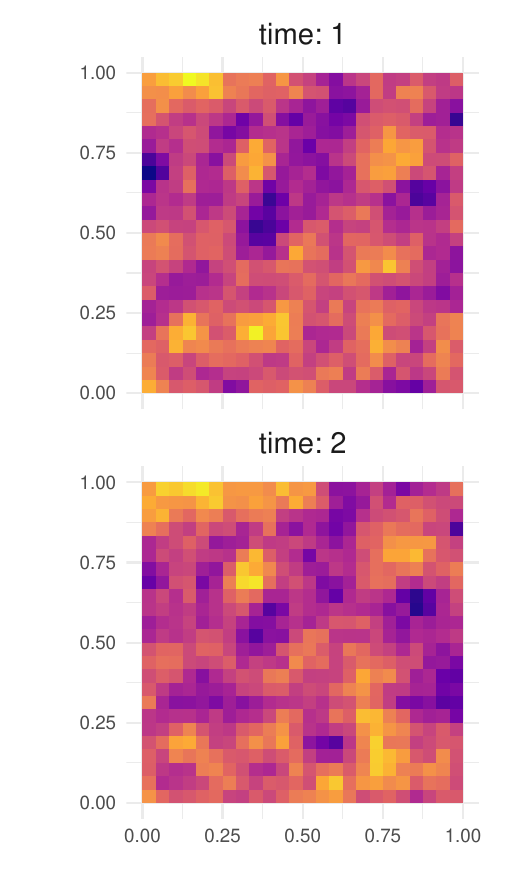}
    \caption{$W(\mathbf{s},t)$}
    \label{fig:app-sep-s-1}
  \end{subfigure}
  \hfill
  \begin{subfigure}[c]{0.31\textwidth}
    \includegraphics[width=0.9\linewidth]{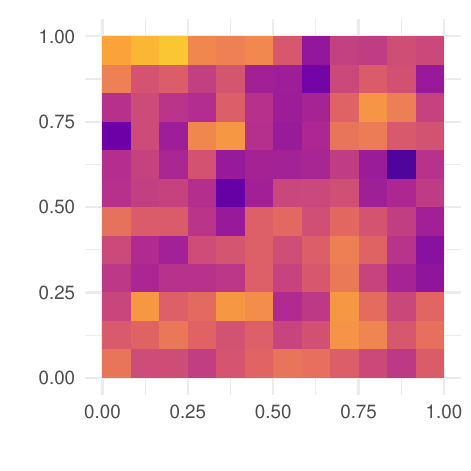}
    \caption{$W(R_{ij})$}
    \label{fig:app-sep-s-2}
  \end{subfigure}
  \hfill
  \begin{subfigure}[c]{0.31\textwidth}
    \includegraphics[width=0.9\linewidth]{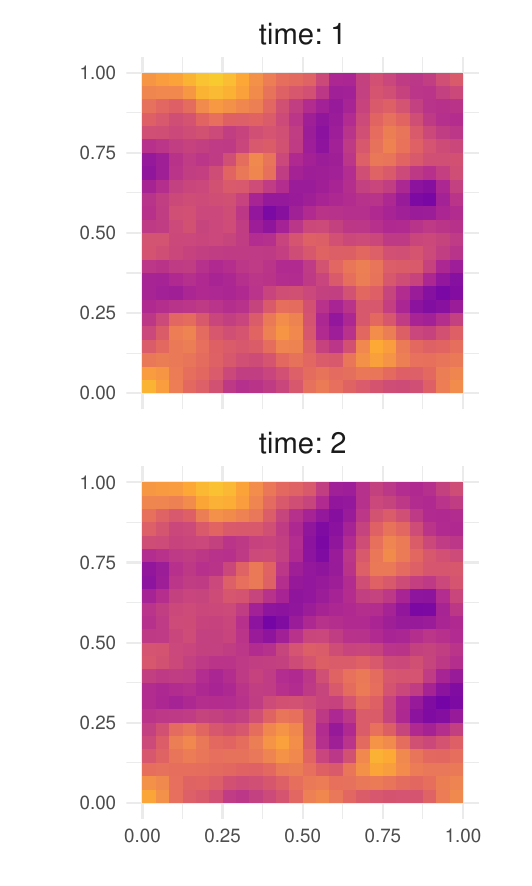}
    \caption{$\widehat{W}(\mathbf{s},t)$}
    \label{fig:app-sep-s-3}
  \end{subfigure}
  \caption{(a) Realization of the continuous process $W(\mathbf{s},t)$ in the \textit{separable} setting with \textbf{strong temporal autocorrelation}, (b) corresponding aggregated observations $W(R_{ij})$ for spatial factor $s_f =2$ and temporal factor $t_f=2$, and (c) the reconstructed process from the disaggregation model $\widehat{W}(\mathbf{s},t)$.}
  \label{fig:appendix-sep-3}
\end{figure}

\begin{figure}[h!]
  \centering
  \begin{subfigure}[c]{0.31\textwidth}
    \includegraphics[width=0.9\linewidth]{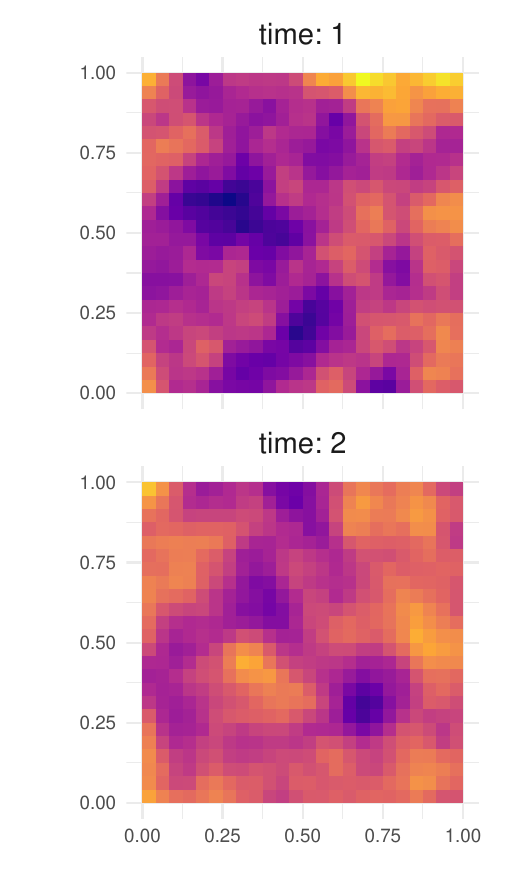}
    \caption{$W(\mathbf{s},t)$}
    \label{fig:ex2-cont}
  \end{subfigure}
  \hfill
  \begin{subfigure}[c]{0.31\textwidth}
    \includegraphics[width=0.9\linewidth]{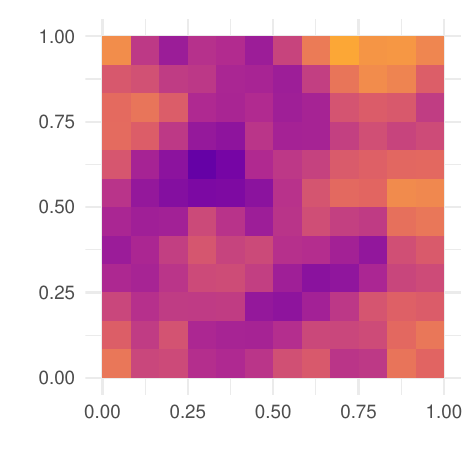}
    \caption{$W(R_{ij})$}
    \label{fig:ex2-agg}
  \end{subfigure}
  \hfill
  \begin{subfigure}[c]{0.31\textwidth}
    \includegraphics[width=0.9\linewidth]{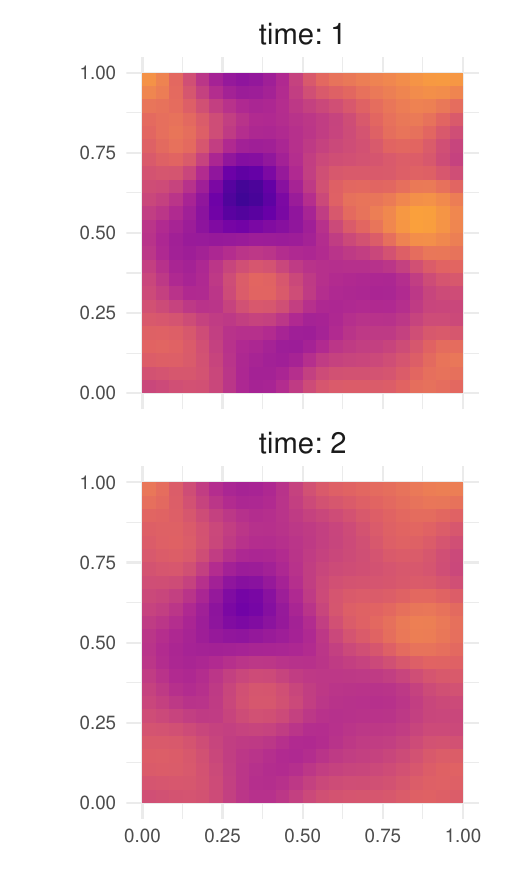}
    \caption{$\widehat{W}(\mathbf{s},t)$}
    \label{fig:ex2-disagg}
  \end{subfigure}
  \caption{(a) Realization of the continuous process $W(\mathbf{s},t)$ in the \textit{non--separable} setting with \textbf{weak temporal autocorrelation}, (b) corresponding aggregated observations $W(R_{ij})$ for spatial factor $s_f =2$ and temporal factor $t_f=2$, and (c) the reconstructed process from the disaggregation model $\widehat{W}(\mathbf{s},t)$.}
  \label{fig:appendix-non-sep-1}
\end{figure}

\begin{figure}[h!]
  \centering
  \begin{subfigure}[c]{0.31\textwidth}
    \includegraphics[width=0.9\linewidth]{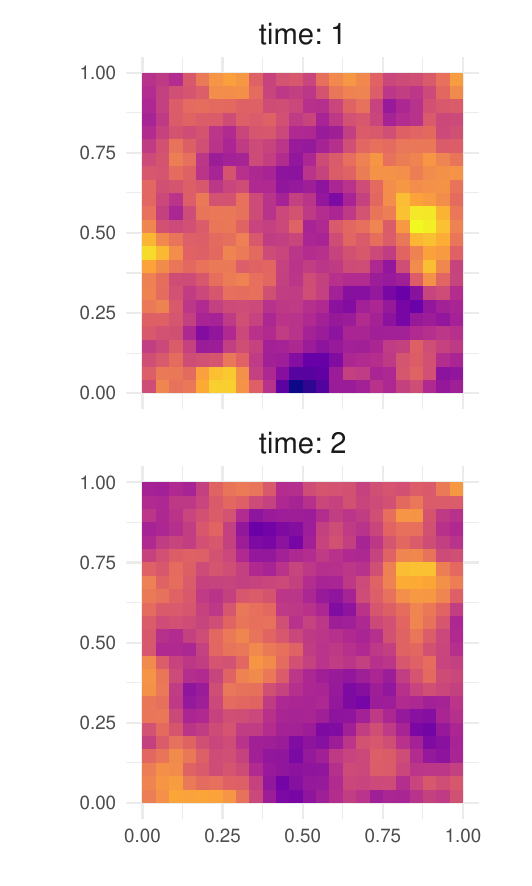}
    \caption{$W(\mathbf{s},t)$}
    \label{fig:ex2-cont}
  \end{subfigure}
  \hfill
  \begin{subfigure}[c]{0.31\textwidth}
    \includegraphics[width=0.9\linewidth]{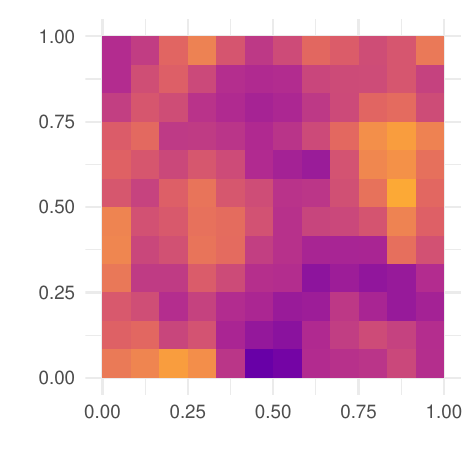}
    \caption{$W(R_{ij})$}
    \label{fig:ex2-agg}
  \end{subfigure}
  \hfill
  \begin{subfigure}[c]{0.31\textwidth}
    \includegraphics[width=0.9\linewidth]{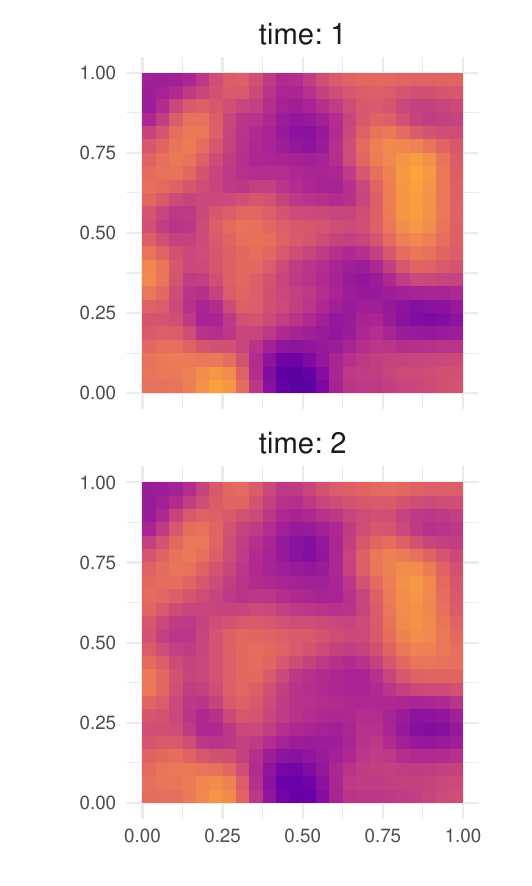}
    \caption{$\widehat{W}(\mathbf{s},t)$}
    \label{fig:ex2-disagg}
  \end{subfigure}
  \caption{(a) Realization of the continuous process $W(\mathbf{s},t)$ in the \textit{non--separable} setting with \textbf{moderate temporal autocorrelation}, (b) corresponding aggregated observations $W(R_{ij})$ for spatial factor $s_f =2$ and temporal factor $t_f=2$, and (c) the reconstructed process from the disaggregation model $\widehat{W}(\mathbf{s},t)$.}
  \label{fig:appendix-non-sep-2}
\end{figure}

\begin{figure}[h!]
  \centering
  \begin{subfigure}[c]{0.31\textwidth}
    \includegraphics[width=0.9\linewidth]{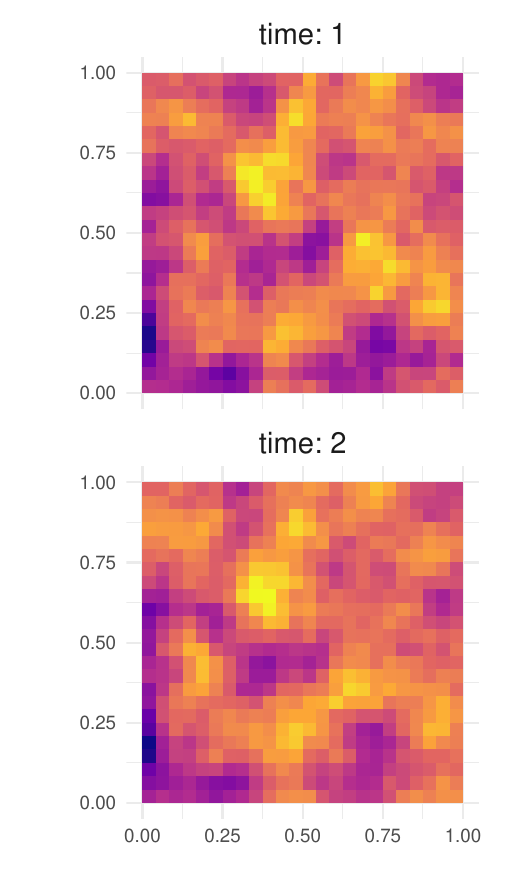}
    \caption{$W(\mathbf{s},t)$}
    \label{fig:ex2-cont}
  \end{subfigure}
  \hfill
  \begin{subfigure}[c]{0.31\textwidth}
    \includegraphics[width=0.9\linewidth]{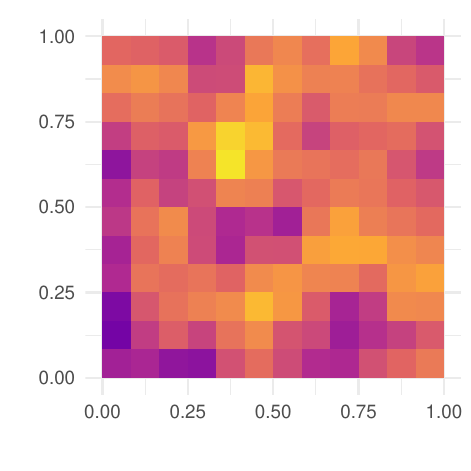}
    \caption{$W(R_{ij})$}
    \label{fig:ex2-agg}
  \end{subfigure}
  \hfill
  \begin{subfigure}[c]{0.31\textwidth}
    \includegraphics[width=0.9\linewidth]{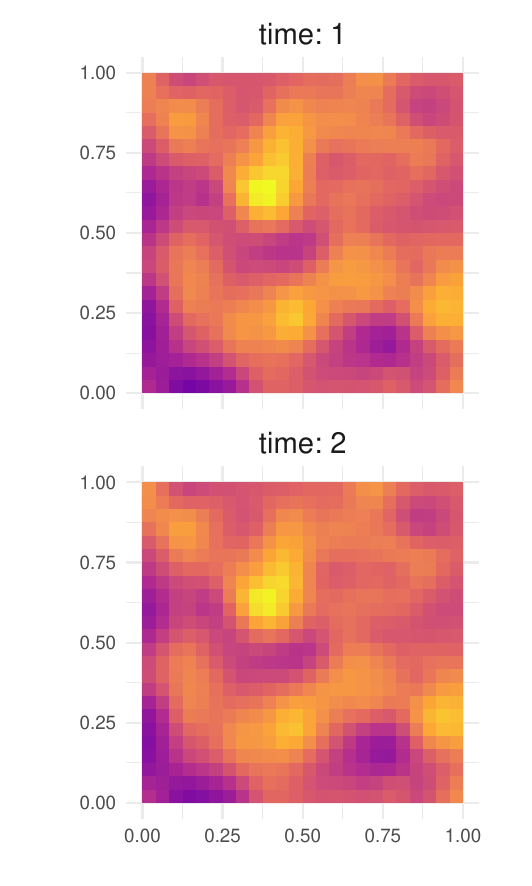}
    \caption{$\widehat{W}(\mathbf{s},t)$}
    \label{fig:ex2-disagg}
  \end{subfigure}
  \caption{(a) Realization of the continuous process $W(\mathbf{s},t)$ in the \textit{non--separable} setting with \textbf{strong temporal autocorrelation}, (b) corresponding aggregated observations $W(R_{ij})$ for spatial factor $s_f =2$ and temporal factor $t_f=2$, and (c) the reconstructed process from the disaggregation model $\widehat{W}(\mathbf{s},t)$.}
  \label{fig:appendix-non-sep-3}
\end{figure}

In this appendix, we provide some realization of the simulation results described in Section~\ref{sec:sim}. Figures \ref{fig:appendix-sep-1} $-$ \ref{fig:appendix-sep-3} show the separable model outcomes for weak, moderate, and strong temporal autocorrelation, while Figures~\ref{fig:appendix-non-sep-1} $-$ \ref{fig:appendix-non-sep-3} present the corresponding results for the non--separable model. All results correspond to the scenario with spatial factor $s_f = 2$ and temporal factor $t_f = 2$, where each pixel of the aggregated data represents the average of $2 \times 2 = 4$ spatial locations and two temporal points, as summarized in Table~\ref{tab:agg-summary}.

\clearpage

\end{appendices}

\end{document}